\documentclass[superscriptaddress,notitlepage,nofootinbib,raggedbottom]{revtex4-2}
\usepackage{natbib}
\usepackage{physics}            
\usepackage{amsmath, amsfonts}
\usepackage{graphicx}
\usepackage[caption=false]{subfig}
\usepackage[english]{babel}
\usepackage[utf8]{inputenc}
\usepackage[autostyle]{csquotes}
\MakeOuterQuote{"}
\usepackage[shortlabels]{enumitem}
\usepackage{dsfont}
\usepackage[hidelinks]{hyperref}	
\usepackage[capitalize]{cleveref}
\usepackage{tcolorbox}
\usepackage{bm}

\newcommand{\set}[1]{\{#1\}}

\makeatletter\AtBeginDocument{\let\@elt\relax}\makeatother
\begin{document}
\title{\texorpdfstring{Finite-key analysis of loss-tolerant quantum key distribution \\ based on random sampling theory}{Finite-key analysis of loss-tolerant quantum key distribution based on random sampling theory}}
\author{Guillermo Currás-Lorenzo}
\email{g.j.curraslorenzo@leeds.ac.uk}
\affiliation{School of Electronic and Electrical Engineering, University of Leeds, Leeds, United Kingdom}
\author{Álvaro Navarrete}
\author{Margarida Pereira}
\affiliation{Escuela de Ingenier$\acute{\textit{\i}}$a de Telecomunicaci$\acute{o}$n, Department of Signal Theory and Communications, University of Vigo, Vigo E-36310, Spain}
\author{Kiyoshi Tamaki}
\affiliation{Faculty of Engineering, University of Toyama, Gofuku 3190, Toyama 930-8555, Japan}
\date{\today}

\begin{abstract}
The core of security proofs of quantum key distribution (QKD) is the estimation of a parameter that determines the amount of privacy amplification that the users need to apply in order to distill a secret key. To estimate this parameter using the observed data, one needs to apply concentration inequalities, such as random sampling theory or Azuma's inequality. The latter can be straightforwardly employed in a wider class of QKD protocols, including those that do not rely on mutually unbiased encoding bases, such as the loss-tolerant (LT) protocol. However, when applied to real-life finite-length QKD experiments, Azuma's inequality typically results in substantially lower secret-key rates. Here, we propose an alternative security analysis of the LT protocol against general attacks, for both its prepare-and-measure and measure-device-independent versions, that is based on random sampling theory. Consequently, our security proof provides considerably higher secret-key rates than the previous finite-key analysis based on Azuma's inequality. This work opens up the possibility of using random sampling theory to provide alternative security proofs for other QKD protocols.
\end{abstract}

\maketitle

Quantum key distribution (QKD) allows two distant users, Alice and Bob, to generate a shared secret key in the presence of an eavesdropper, Eve, with unbounded computational power \cite{scarani,lo}. To prove the security of QKD, we often consider the error rate that Alice and Bob would have obtained in a fictitious scenario,  known as the phase-error rate, which directly bounds the amount of sifted-key information that could have leaked to Eve, and determines the amount of privacy amplification that the users need to apply to distill a secret key \cite{koashi,koashi2,hayashi2012concise,tsurumaru2020leftover}. Since Alice and Bob cannot directly observe the phase-error rate, they must estimate it using the data collected in the test rounds, i.e.\ the detected rounds which are not used to generate the sifted key. For this estimation, it is indispensable to employ statistical techniques. For example, in the case of the BB84 protocol \cite{bennett} without source flaws, one can use the fact that Alice's encoding bases are mutually unbiased to estimate the $Z$-basis phase-error rate from the $X$-basis bit-error rate, and vice-versa, using random sampling theory \cite{fung2010practical,tomamichel2012tight}. In protocols where the users do not employ two mutually unbiased bases, the detection statistics of a particular round may depend on the basis choices made in previous rounds, and Azuma's inequality \cite{azuma} has been typically applied to deal with this dependency \cite{boileau2005unconditional,tamaki2014loss,mizutani2,mizutani2019quantum}. However, recently, Maeda \textit{et al.\ }\cite{maeda2019repeaterless} have successfully applied a non-trivial security analysis based on random sampling theory to a twin-field QKD variant in which the users do not employ mutually unbiased encoding bases. This work raises the obvious question of whether random sampling theory could also be applied to other protocols that do not use mutually unbiased bases, and whose security proofs currently rely on Azuma's inequality. Since the estimation of Eve's side information is the core of QKD security proofs, investigating the possibility of using different estimation techniques deepens our understanding of QKD protocols and their security. Moreover, it has important experimental implications, in terms of the secret-key rate obtainable, since concentration bounds for \textit{independent} random variables, such as the Chernoff bound, are typically tighter than those for \textit{dependent} random variables, such as Azuma's inequality.

One obvious candidate to investigate is the loss-tolerant (LT) protocol \cite{tamaki2014loss}, a three state protocol that is resistant to state preparation flaws (SPFs), which arise from the finite precision of modulation devices. Earlier attempts to address SPFs \cite{gottesman} resulted in a performance that degraded very quickly with moderate-to-high channel losses. Conversely, even in the presence of large SPFs and high losses, the performance of the LT protocol is close to that of a perfect four-state BB84 protocol, at least in the limit of infinitely-long keys \cite{tamaki2014loss}. Recent works \cite{pereira,pereira2,navarrete2021hopefully} have shown that one can prove the security of the LT protocol in the presence of additional source imperfections, such as mode dependencies, Trojan horse attacks or pulse correlations, as long as one can ensure that their magnitude is sufficiently small. Also, the LT protocol can be combined with measurement-device-independent (MDI) QKD \cite{lo2012measurement} to guarantee the security in the presence of arbitrarily flawed detectors. Moreover, the LT protocol is highly practical and can be implemented with off-the-shelf devices. In fact, several experiments have implemented the LT protocol \cite{xu2015experimental,tang2016experimental}, and a variation of it \cite{boaron2018secure} set a fibre QKD distance record. For these reasons, a deep understanding of its security is of theoretical and practical interest. 

Clearly, the LT protocol does not rely on using two mutually encoding bases. For starters, in its standard three-state formulation, Alice only emits one of the two $X$-basis states. However, even if one were to apply the LT idea to a four-state protocol, the encoding bases would still not be mutually unbiased, due to the SPFs. Thus, Azuma's inequality has been used in both the asymptotic \cite{tamaki2014loss} and finite-key \cite{mizutani2} security proofs of the LT protocol. In the asymptotic regime, the specific statistical technique employed does not affect the performance, since the deviation terms vanish in the limit of infinitely-long keys. However, choosing the tightest statistical technique available does have an impact on the key rate obtainable in (existing and future) real-life finite-length implementations of the LT protocol.

In this paper, we show how the finite-key security of the LT protocol against general attacks can be reduced to a random sampling problem, for both its original prepare-and-measure (P\&M) version and its MDI version. This random sampling problem can be solved using concentration inequalities for sums of independent random variables, which results in tighter bounds than those of a previous analysis \cite{mizutani2} based on Azuma's inequality. Our paper is structured as follows. In \cref{sec:general}, we present our general statistical analysis, inspired by that of Ref.~\cite{maeda2019repeaterless}, and apply it to a generic scenario. In \cref{sec:pam}, we show how this analysis can be used to estimate the phase-error rate of the P\&M LT protocol, and in \cref{sec:mdi}, we do the same for the MDI LT protocol. In \cref{sec:security_bounds}, we give an expression for the secret-key rate obtainable in both protocols. In \cref{sec:numerical_results}, we simulate the secret-key rate obtainable for different values of the block size, and compare it with that of alternative analyses. Finally, in \cref{sec:conclusion}, we conclude our paper.

\section{General statistical analysis}\label{sec:general}
In this section, we present our general estimation procedure and apply it to a generic scenario, which we denote as the Tagged Virtual Protocol (TVP). Its name refers to the fact that, as we will see in \cref{sec:pam,sec:mdi}, one can draw an equivalence between the TVP and the virtual protocols of both LT P\&M QKD and LT MDI QKD, once the users probabilistically assign tags to their emissions.

In the TVP, the users emit, amongst others, the states $\rho_{\rm vir}$, $\rho_{\rm pos}$ and $\rho_{\rm neg}$, with probabilities $p_{\rm vir}$, $p_{\rm pos}$ and $p_{\rm neg}$. These may be states sent by Alice, in the P\&M protocol, or joint states sent by Alice and Bob, in the MDI protocol. Also, $\rho_{\rm vir}$ is one of the virtual states, emitted only in the virtual protocol, while $\rho_{\rm pos}$ and $\rho_{\rm neg}$ are actual states, emitted also in the actual protocol. These states satisfy
\begin{equation}
\label{eq:rho_vir}
\rho_{\rm vir} = c_{\rm pos} \rho_{\rm pos} -  c_{\rm neg} \rho_{\rm neg}
\end{equation}
where $c_{\rm pos}$ and $c_{\rm neg}$ are some non-negative coefficients such that $c_{\rm pos} - c_{\rm neg} = 1$. For reasons that will become clear later on, we assume that the users assign a tag of $t \in \{\rm vir, pos, neg\}$ to each emission of $\rho_t$. That is, each emission of $\rho_{\rm vir}$ is trivially assigned a tag $t = \textrm{vir}$, and so on. In the quantum communication phase of the protocol, some of these emissions will be detected. Here, a "detection" refers to any process that depends on Eve's attack and distinguishes some emissions from others. For the P\&M protocol, we will define a detection as an event in which Bob obtained a particular measurement result, and for the MDI protocol, as an event in which Charlie reports a projection to a particular Bell state. We denote by $N_{t}$ the number of detected emissions with a tag of $t$, i.e., the number of detected emissions of $\rho_t$. In the actual protocol, the outcome of the random variables $N_{\rm pos}$ and $N_{\rm neg}$ can be directly observed by the users, but the outcome of $N_{\rm vir}$ cannot, and must be estimated. Thus, the objective of the analysis is to find a statistical relationship between $N_{\rm vir}$, $N_{\rm pos}$ and $N_{\rm neg}$; more specifically, we want to find a function $f$ such that $\Pr[N_{\rm vir} > f(N_{\rm pos}, N_{\rm neg};\varepsilon)] \leq \varepsilon$, where $\varepsilon$ can be made arbitrarily small.

The starting point of the analysis is \cref{eq:rho_vir}, which we now rewrite as
\begin{equation}
\label{eq:rho_plus}
\rho_{\rm pos} = p_{\rho_{\rm vir} \vert \rm pos} \rho_{\rm vir}  + p_{\rho_{\rm neg} \vert \rm pos} \rho_{\rm neg},
\end{equation}
where $p_{\rho_{\rm vir} \vert \rm pos} = 1/c_{\rm pos}$ and $p_{\rho_{\rm neg}|\rm pos} = c_{\rm neg}/c_{\rm pos}$. \Cref{eq:rho_plus} implies that sending $\rho_{\rm pos}$ is equivalent to sending $\rho_{\rm vir}$ with probability $p_{\rho_{\rm vir} \vert \rm pos}$ and $\rho_{\rm neg}$ with probability $p_{\rho_{\rm neg} \vert \rm pos}$. That is, the TVP is indistinguishable from the following scenario:
\vspace{-4pt}
\begin{itemize}
\itemsep-0.33em
    \item[--] The users select tag $t \in \{\rm vir, pos, neg\}$ with probability $p_t$.
    \item[--] If $t = {\rm pos}$, the users emit $\rho_{\rm vir}$ with probability $p_{\rho_{\rm vir} \vert \rm pos}$, or $\rho_{\rm neg}$ with probability $p_{\rho_{\rm neg} \vert \rm pos}$.
    \item[--]  If $t \in \{\rm vir, neg\}$, the users emit $\rho_t$.
\end{itemize}
\vspace{-4pt}
In the above scenario, some emissions of $\rho_{\rm vir}$ will have a tag of ${\rm ``vir"}$, and some will have a tag of ${\rm ``pos"}$, but they are otherwise identical. The same is true for emissions of $\rho_{\rm neg}$ with tags of ${\rm ``neg"}$ and ${\rm ``pos"}$. Thus, one can go even further, and think of another equivalent scenario in which the users first decide the quantum state that they emit, and then probabilistically assign a tag to it. Namely:
\vspace{10pt}
\begin{tcolorbox}
\textbf{\textit{Modified scenario}}
\begin{itemize}
    \item The users select and emit the state $\rho_{x} \,\in \{\rho_{\rm vir},\rho_{\rm neg}\}$ with probability $\tilde{p}_{\rho_{x}} := p_{x} + p_{\rm pos}  p_{\rho_{x} \vert \rm pos}$.
    \item Next, they assign their emission the tag $t=x$ with probability $\tilde{p}_{x | \rho_{x}} := p_{x}/\tilde{p}_{\rho_{x}}$, or the tag $t = {\rm pos}$ with probability $\tilde{p}_{{\rm pos} | \rho_{x}} := 1 - \tilde{p}_{x | \rho_{x}}$.
\end{itemize}
\end{tcolorbox}
\vspace{10pt}

This modified scenario is equivalent to the TVP in terms of tags, because:
\begin{enumerate}
    \item The overall probability to assign a particular tag $t \in \{\rm vir, pos, neg\}$ is the same in both scenarios, i.e.\ $p_t$.
    \item The quantum state emitted given a particular tag $t$ is the same in both scenarios, i.e.\ $\rho_t$.
\end{enumerate}

\begin{figure}[ht]
		\centering
		\includegraphics[width=0.6\textwidth]{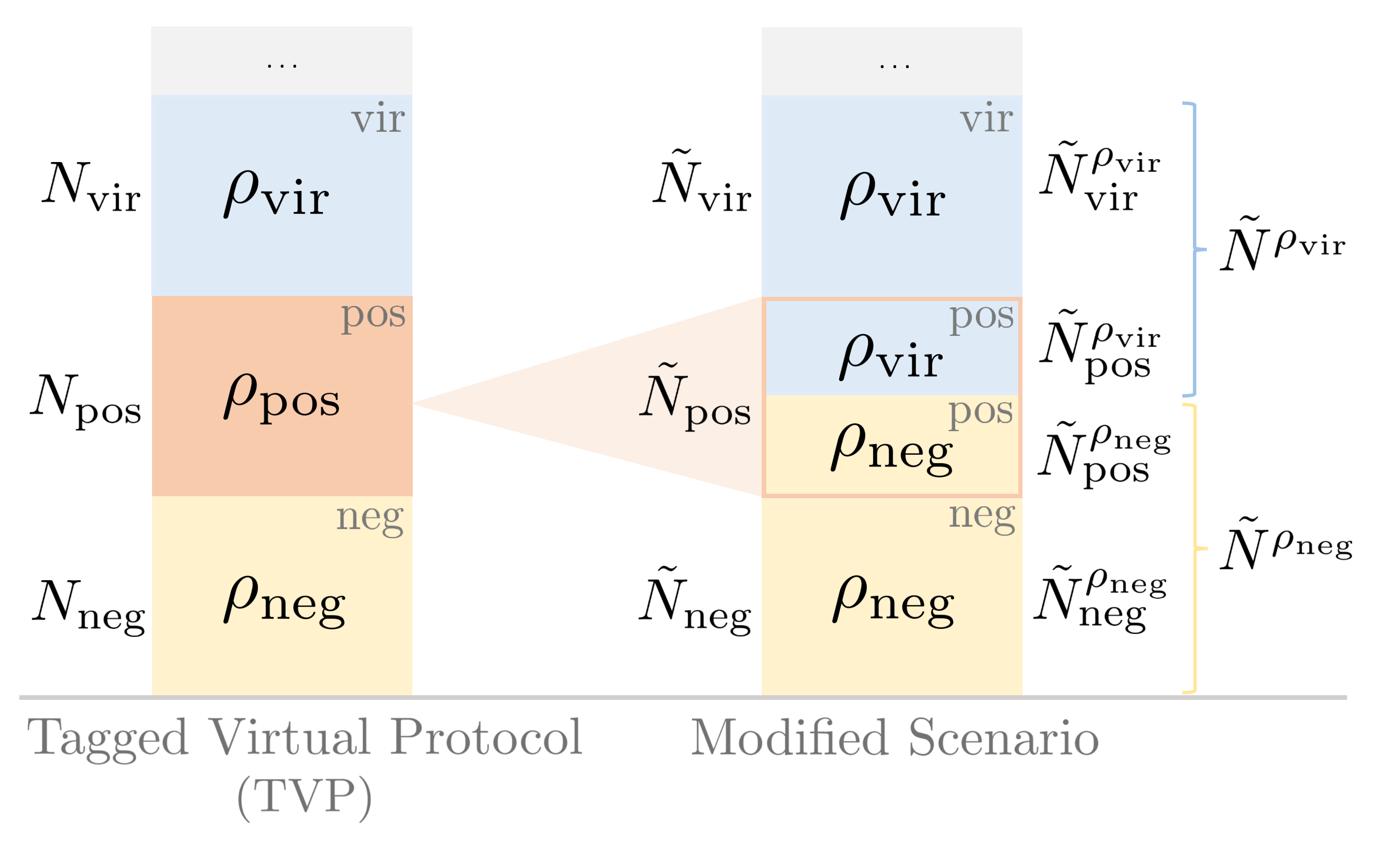}
		\caption{Relationship between the Tagged Virtual Protocol (TVP) and the modified scenario. In the modified scenario, each emission of $\rho_{\rm vir}$ ($\rho_{\rm neg}$) is assigned a tag of either "vir" ("neg") or "pos" with a fixed probability, in such a way that emissions with a tag of $t \in \{\rm vir, neg, pos\}$ are equivalent to emissions of $\rho_t$ in the TVP. In the modified scenario, the detection statistics of each emission must be independent of the tag assigned to it, since Eve does not have any tag information. Hence, each of the $\tilde{N}^{\rho_{\rm vir}}$ ($\tilde{N}^{\rho_{\rm neg}}$) detected emissions of $\rho_{\rm vir}$ ($\rho_{\rm neg}$) is assigned a tag of either "vir" ("neg") or "pos" with the \textit{a priori} fixed probability. This allows us to find a statistical relationship between the random variables $\tilde N_{\rm vir} := \tilde N_{\rm vir}^{\rho_{\rm vir}}$, $\tilde N_{\rm pos} := \tilde N_{\rm pos}^{\rho_{\rm vir}} + \tilde N_{\rm pos}^{\rho_{\rm neg}}$ and $\tilde{N}_{\rm neg} := \tilde N_{\rm neg}^{\rho_{\rm neg}}$ using a random sampling analysis, see \cref{eq:Nvir3}. Since the TVP is equivalent to the modified scenario, the same relationship must hold for the random variables $N_{\rm vir}$, $N_{\rm pos}$ and $N_{\rm neg}$ in the TVP, see \cref{eq:Nvir4}.} \label{fig:diagram}
	\end{figure}

In the modified scenario, let $\tilde{N}^{\rho_x}_{t}$ be the number of detected emissions of $\rho_{x}$ with a tag of $t$, $\tilde{N}^{\rho_{x}} = \sum_t \tilde{N}^{\rho_x}_{t}$ be the total number of detected emissions of $\rho_{x}$, and $\tilde{N}_{t} = \sum_x \tilde{N}^{\rho_x}_{t}$ be the total number of detected emissions with a tag of $t$. That is, $\tilde N_{\rm vir} = \tilde N_{\rm vir}^{\rho_{\rm vir}}$, $\tilde N_{\rm pos} = \tilde N_{\rm pos}^{\rho_{\rm vir}} + \tilde N_{\rm pos}^{\rho_{\rm neg}}$, and $\tilde{N}_{\rm neg} = \tilde N_{\rm neg}^{\rho_{\rm neg}}$. The equivalence above implies that, for any attack by Eve, the set of random variables $\{N_{\rm vir}, N_{\rm pos},N_{\rm neg}\}$ in the TVP has an identical distribution as the set $\{\tilde{N}_{\rm vir}, \tilde{N}_{\rm pos},\tilde{N}_{\rm neg}\}$ in the modified scenario. Hence, if we find a function $f$ such that $\Pr\big[\tilde N_{\rm vir} > f(\tilde N_{\rm pos}, \tilde N_{\rm neg};\varepsilon)\big] \leq \varepsilon$ in an execution of the modified scenario, then it must also be the case that $\Pr\big[N_{\rm vir} > f(N_{\rm pos}, N_{\rm neg};\varepsilon)\big] \leq \varepsilon$ in an execution of the TVP. The equivalence between the two scenarios is shown in \cref{fig:diagram}.
	
The random tag assignments in the modified scenario allow us to find a bound on $\tilde N_{\rm vir}$ by using a random sampling analysis. The key idea is that the probability to assign a particular tag to a particular emission must be independent of whether the emission is detected or not, since the tag assignment does not change the emitted quantum state, and Eve does not have any tag information. Thus, each of the $\tilde{N}^{\rho_{\rm vir}}$ detected emissions of $\rho_{\rm vir}$ is  assigned a random tag of ${\rm ``vir"}$ or ${\rm ``pos"}$ with probabilities $\tilde{p}_{{\rm vir} | \rho_{\rm vir}}$ and $\tilde{p}_{{\rm pos} | \rho_{\rm vir}} = 1- \tilde{p}_{{\rm vir} | \rho_{\rm vir}}$, respectively. This implies that $\tilde{N}^{\rho_{\rm vir}}_{\rm vir}$ is a random sample of a population of $\tilde{N}^{\rho_{\rm vir}} = \tilde{N}^{\rho_{\rm vir}}_{\rm vir} + \tilde{N}^{\rho_{\rm vir}}_{\rm pos}$ elements, where each item is sampled with probability $\tilde{p}_{{\rm vir} | \rho_{\rm vir}}$. In Appendix A, we show that this implies that, except with probability $\varepsilon/2$,
\begin{equation}
\label{eq:Nvir}
    \tilde{N}^{\rho_{\rm vir}}_{\rm vir} \leq g_U \left(\tilde N^{\rho_{\rm vir}}_{\rm  pos}, \tilde{p}_{{\rm vir} | \rho_{\rm vir}}, \varepsilon/2  \right),
\end{equation}
where $g_U$ is defined in \cref{eq:functions_random_sampling}.
Similarly $\tilde N^{\rho_{\rm neg}}_{\rm pos}$ is the size of a random sample of a population of $\tilde N^{\rho_{\rm neg}} =  \tilde N^{\rho_{\rm neg}}_{\rm pos} +  \tilde N^{\rho_{\rm neg}}_{\rm neg}$ elements, where each item is sampled with probability $\tilde{p}_{{\rm pos} | \rho_{\rm neg}}$. This implies that, except with probability $\varepsilon/2$,
\begin{equation}
\label{eq:Nnegpos}
	\tilde N^{\rho_{\rm neg}}_{\rm pos} \geq g_L\left(\tilde N_{\rm neg},\tilde{p}_{{\rm pos} | \rho_{\rm neg}},\varepsilon/2 \right),
\end{equation}
where $g_L$ is defined in \cref{eq:functions_random_sampling}.

Using the relations $\tilde N_{\rm vir} = \tilde N_{\rm vir}^{\rho_{\rm vir}}$, $\tilde N_{\rm pos} = \tilde N_{\rm pos}^{\rho_{\rm vir}} + \tilde N_{\rm pos}^{\rho_{\rm neg}}$, and $\tilde{N}_{\rm neg} = \tilde N_{\rm neg}^{\rho_{\rm neg}}$, in combination with \cref{eq:Nvir,eq:Nnegpos}, we have that
\begin{equation}
\label{eq:Nvir3}
\begin{aligned}
    \tilde N_{\rm vir} &\leq g_U \left(\tilde N_{\rm pos} - \tilde N_{\rm pos}^{\rho_{\rm neg}}, \tilde{p}_{{\rm vir} | \rho_{\rm vir}}, \varepsilon/2  \right) \\
     &\leq g_U \left(\tilde N_{\rm pos} - g_L\left(\tilde N_{\rm neg},\tilde{p}_{{\rm pos} | \rho_{\rm neg}},\varepsilon/2 \right),\tilde{p}_{{\rm vir} | \rho_{\rm vir}}, \varepsilon/2  \right),
\end{aligned}
\end{equation}
except with probability $\varepsilon$, where in the first inequality we have used \cref{eq:Nvir}, and in the second inequality we have used \cref{eq:Nnegpos} and the fact that $g_U$ is an increasing function with respect to its first argument.

As explained above, the random variables $\{N_{\rm vir}, N_{\rm pos},N_{\rm neg}\}$ in the TVP are identically distributed as the random variables $\{\tilde{N}_{\rm vir}, \tilde{N}_{\rm pos},\tilde{N}_{\rm neg}\}$ in the modified scenario. Thus, \cref{eq:Nvir3} implies that, in the virtual protocol
\begin{equation}
\label{eq:Nvir4}
    N_{\rm vir} \leq g_U \left(N_{\rm pos} - g_L\Big(N_{\rm neg},\tilde{p}_{{\rm pos} | \rho_{\rm neg}},\varepsilon/2 \Big), \tilde{p}_{{\rm vir} | \rho_{\rm vir}}, \varepsilon/2 \right) := f(N_{\rm pos}, N_{\rm neg};\varepsilon),
\end{equation}
except with probability $\varepsilon$, as required. Since $N_{\rm pos}$ and $N_{\rm neg}$ are observables of the actual protocol, Alice and Bob can use their observed values to obtain an upper bound on $N_{\rm vir}$.

In \cref{sec:pam,sec:mdi}, we explain how to apply this statistical analysis to the LT protocol, for both its P\&M and MDI versions. In this protocol, the virtual states and the actual states are all in the same qubit space.
Because of this, each virtual state can be expressed as an operator-form linear function of the actual states. However, this linear function does not necessarily have one positive term and one negative term, as in \cref{eq:rho_vir}. To apply the analysis above, the users will first probabilistically assign tags of ${\rm ``pos"}$ and ${\rm ``neg"}$ to some of their emissions, in such a way that the average state with a tag of $t \in \{\rm pos, neg\}$ is $\rho_t$. After these tag assignments, the resulting \textit{tagged} virtual protocol will be equivalent to the TVP, shown on the left side of \cref{fig:diagram}.
	
\section{Prepare-and-measure protocol}
\label{sec:pam}

In this section, we apply our analysis to the P\&M LT protocol \cite{tamaki2014loss}. For each round, Alice sends Bob a pure state $\ket{\psi_j}_a$ with probability $p_j$, $j \in \{ 0_Z, 1_Z, 0_X\}$, where emissions of $\ket{\psi_{0_X}}_a$ ($\ket{\psi_{0_Z}}_a$ and $\ket{\psi_{1_Z}}_a$) are considered to belong to the $X$ ($Z$) basis. The only assumption needed to apply our analysis is that Alice's states are characterised and linearly dependent, i.e. they are all in the same qubit space. For simplicity, in this discussion we assume that the states are in the $XZ$ plane of the Bloch sphere; in \cref{app:coef}, we show how to apply our results in the general case. Bob measures the incoming signals in the $Z$ or in the $X$ basis, with probabilities $p_{Z_B}$ and $p_{X_B}$, respectively. We do not need to assume that Bob's measurement bases are mutually unbiased, but we do assume that his choice of basis is fully random, and that the detection efficiency is the same for both bases.  Afterwards, Bob reveals which rounds were detected, and both users reveal their basis choice in those rounds. The sifted key is generated from the detected events in which Alice and Bob both chose the $Z$ basis. The detected rounds in which Bob chose the $X$ basis are considered to be test rounds. In these, Bob will reveal his measurement result. The full protocol description is given in \cref{app:protocol_description_pm}.

The objective of the security analysis is to estimate the number of phase errors in the sifted key, using the test data. To define this quantity, we consider an equivalent entanglement-based virtual protocol, in which Alice replaces the key emissions by the generation of the entangled state
\begin{align}
  &\ket{\Psi_Z}_{Aa} = \frac{1}{\sqrt{2}} \big(\ket{0_Z}_A \ket{\psi_{0_Z}}_a +  \ket{1_Z}_A \ket{\psi_{1_Z}}_a\big), \label{eq:z_state}
\end{align}
where $a$ is the photonic system sent to Bob and $A$ is Alice's fictitious qubit ancilla, which she keeps in her lab. For simplicity, in \cref{eq:z_state}, we have assumed that $p_{0_Z} = p_{1_Z}$.
The key generated in the actual protocol is equivalent to the key that Alice and Bob would obtain by performing a $Z$-basis measurement on the systems $A$ and $a$ of the detected rounds in which Alice generated $\ket{\Psi_Z}_{Aa}$. 
The number of phase errors is defined as the number of errors that Alice and Bob would have observed if they had measured these systems $A$ and $a$ in the $X$ basis instead. This is equivalent to a scenario in which, in the key rounds, Alice sends Bob the virtual states
\begin{align}
    \ket{\psi_{\rm vir_\alpha}}_a = \frac{\ket{\psi_{0_Z}}_a + (-1)^{\alpha} \ket{\psi_{1z}}_a}{\sqrt{2(1-(-1)^{\alpha} \braket{\psi_{0_Z}}{\psi_{1_Z}}_a)}},
    \label{eq:virtual}
\end{align}
with probabilities 
\begin{align}
    p_{\rm vir_\alpha} = \frac{1}{2} p_{Z_A} (1 - (-1)^\alpha \braket{\psi_{0_Z}}{\psi_{1_Z}}_a),
\end{align}
and Bob measures these states in the $X$ basis. Here, $p_{Z_A}$ is the probability that Alice selects the $Z$ basis, and $\alpha \in \{0,1\}$. Thus, Alice's choice of state in the virtual protocol can be equivalently described by assuming that she fictitiously prepares the entangled state 
\begin{equation}
\begin{aligned}
    \ket{\Psi_{\rm vir}}_{Sa} =& \sqrt{p_{\rm vir_0} p_{Z_B}} \ket{0}_S \ket{\psi_{\rm vir_0}}_a + \sqrt{p_{\rm vir_1}p_{Z_B}} \ket{1}_S \ket{\psi_{\rm vir_1}}_a + \sqrt{p_{0_Z}p_{X_B}} \ket{2}_S \ket{\psi_{0_Z}}_a  \\ +& \sqrt{p_{1_Z}p_{X_B}} \ket{3}_S \ket{\psi_{1_Z}}_a + \sqrt{p_{0_X}p_{X_B}} \ket{4}_S \ket{\psi_{0_X}}_a + \sqrt{p_{0_X}p_{Z_B}} \ket{5}_S \ket{\psi_{0_X}}_a,
    \label{eq:vir_prot}
\end{aligned}
\end{equation}
and then performs a measurement on system $S$. Note that $S$ holds information about Alice's and Bob's setting choices. For instance, $\ket{2}_S$ represents the events in which Alice selects the actual state $\ket{\psi_{0_Z}}_{a}$ and Bob chooses the $X$ basis. In the right-hand side of Eq.~(\ref{eq:vir_prot}), the first two terms are associated with virtual events. That is, the events in which Alice and Bob select the $Z$ basis in the actual protocol, but their basis choice is replaced by the $X$ basis in the virtual protocol. All the other terms in Eq.~(\ref{eq:vir_prot}) correspond to actual events that occur in the actual protocol.

In the virtual protocol that we have just defined, the occurrence of a phase error is defined as an event in which Alice measures system $S$, obtains the outcome 0 (1), and Bob's $X$-basis measurement outputs the bit value 1 (0). The measurement statistics associated with these events cannot be directly observed, since the virtual states are never sent in the actual protocol. However, as we show in \cref{app:coef}, one can exploit the fact that the virtual states and the actual states live in the same qubit space to find an operator-form linear relationship between the virtual states and the actual states. Namely,
\begin{align}
    &\rho_{\rm vir_0}  = c_{0_Z \vert \rm vir_0} \rho_{0_Z} + c_{1_Z \vert \rm vir_0} \rho_{1_Z} + c_{0_X \vert \rm vir_0} \rho_{0_X}, \nonumber \\
    &\rho_{\rm vir_1} = c_{0_Z \vert \rm vir_1} \rho_{0_Z} + c_{1_Z \vert \rm vir_1} \rho_{1_Z} + c_{0_X \vert \rm vir_1} \rho_{0_X},
    \label{eq:qubit_space}
\end{align}
where $\rho_{\rm vir_\alpha} \equiv \ketbra{\psi_{\rm vir_\alpha}}_a$, $\rho_{j} \equiv \ketbra{\psi_{j}}_a$, and the coefficients $c_{j \vert \rm vir_\alpha}$ can be positive, negative or zero depending on the form of the actual states $\{\ket{\psi_j}_a\}$. For example, when there are no SPFs, the emitted states are $\ket{\psi_{0_Z}}_a = \ket{0_Z}_a$, $\ket{\psi_{1_Z}}_a = \ket{1_Z}_a$ and $\ket{\psi_{0_X}}_a = \ket{0_X}_a$; and \cref{eq:qubit_space} becomes $\rho_{\rm vir_0} = \rho_{0_X}$ and $\rho_{\rm vir_1} = \rho_{0_Z}+\rho_{1_Z}-\rho_{0_X}$. Next, in order to employ the analysis in Section I, we rewrite  Eq.~(\ref{eq:qubit_space}) as
\begin{align}
&\rho_{\rm vir_0} = c_{\rm pos_0} \rho_{\rm pos_0} -  c_{\rm neg_0} \rho_{\rm neg_0}, \label{eq:psivir0posneg} \\
&\rho_{\rm vir_1} = c_{\rm pos_1} \rho_{\rm pos_1} -  c_{\rm neg_1} \rho_{\rm neg_1}, \label{eq:psivir1posneg}
\end{align}
where, for $t \in \{\rm pos,neg\}$ and $\alpha \in \{0,1\}$,
\begin{align}
    &c_{t_\alpha} = \sum_{j \in \mathcal{S}_{t_\alpha}} |c_{j\vert \rm vir_\alpha}|, \\
    &\rho_{t_\alpha} = \sum_{j \in \mathcal{S}_{t_\alpha}} p_{j|t_\alpha} \dyad{\psi_j}_a.
    \label{eq:pos&neg}
\end{align}
In \cref{eq:pos&neg}, $\mathcal{S}_{\rm pos_\alpha}$ ($\mathcal{S}_{\rm neg_\alpha}$) is the set of indices $j$ such that $c_j^{\alpha}$ is positive (negative), and
\begin{align}
    p_{j|t_\alpha} = \frac{|c_{j\vert \rm vir_\alpha}|}{c_{t_\alpha}}.
    \label{eq:prob_ibeta}
\end{align}
Now, each of \cref{eq:psivir0posneg,eq:psivir1posneg} is identical to \cref{eq:rho_vir}, the starting point of the statistical fluctuation analysis introduced in \cref{sec:general}. We will apply this analysis to estimate the detection statistics of each virtual state, separately. Recall that, in the TVP defined in \cref{sec:general}, the states sent are $\rho_{\rm vir}$, $\rho_{\rm pos}$ and $\rho_{\rm neg}$ (see \cref{fig:diagram}). However, n the virtual protocol defined above, Alice does not emit the states $\rho_{\rm pos_0}$, $\rho_{\rm pos_1}$, $\rho_{\rm neg_0}$ and $\rho_{\rm neg_1}$. Instead, Alice will probabilistically assign tags of $t_0 \in \{\rm pos_0,neg_0\}$ and $t_1 \in \{\rm pos_1,neg_1\}$ to some of her emissions, in such a way that the average state with a tag of $t_0$ ($t_1$) is $\rho_{t_0}$ ($\rho_{t_1}$). After doing so, we can draw an equivalence between the virtual protocol and the TVP.

More concretely, let us consider the events in which Alice emits $\ket{\psi_j}_a$, $j \in \{ 0_Z, 1_Z, 0_X\}$, and Bob chooses the $X$ basis, corresponding to measuring system $S$ of \cref{eq:vir_prot} in $2$, $3$ or $4$. Each of these events occurs with probability $p_{j,X_B} = p_j p_{X_B}$, and is assigned a tag of $t_\alpha \in \{{\rm pos_{\alpha}}, {\rm neg_{\alpha}}\}$ with probability
\begin{equation}
\label{eq:tag_probability}
p_{t_{\alpha}|j,X_B} = \frac{p_{t_\alpha} p_{j|t_{\alpha}}}{p_j p_{X_B}},
\end{equation}
or a tag of $t_{\alpha} = {\rm junk_{\alpha}}$ otherwise; where $\alpha \in \{0,1\}$, $p_{j|t_{\alpha}}$ is given by \cref{eq:prob_ibeta}, and $p_{t_{\alpha}}$ is the total probability of assigning tag $t_{\alpha}$. Note that the assignment of tag $t_0$ and of tag $t_1$ is done independently: each of these emissions will have both a tag of $t_0$ and a tag of $t_1$. This is allowed because our key idea relies only on a probabilistic assignment of a tag, and even if multiple assignments are made for a single pulse, the argument still holds. In \cref{eq:tag_probability}, the conditional probabilities $p_{t_{\alpha}|j,X_B}$ become fixed once one chooses the value of $p_{t_\alpha}$, which must be such that $p_{t_\alpha} \leq p_j p_{X_B}/p_{j|t_\alpha}$ for all $j \in \{0_Z, 1_Z, 0_X\}$, since $p_{t_\alpha \vert j, X_B} \leq 1$. In order to waste as few test rounds as possible, and thus obtain a tight estimate of the number of phase errors, we assume that Alice chooses the largest possible value of $p_{t_\alpha}$, given by
\begin{equation}
    p_{t_\alpha} = \min_{j} \frac{p_j p_{X_B}}{p_{j|t_\alpha}}.
    \label{eq:px_condition}
\end{equation}
Moreover, in the virtual protocol, Alice assigns a deterministic tag of $t_0 = {\rm vir_0}$ ($t_1 = {\rm vir_1}$) to each emission of $\ket{\psi_{\rm vir_0}}_a$ ($\ket{\psi_{\rm vir_1}}_a$),  corresponding to $S=0$ ($S=1$).

\begin{figure}[ht]
		\centering	\includegraphics[width=0.6\textwidth]{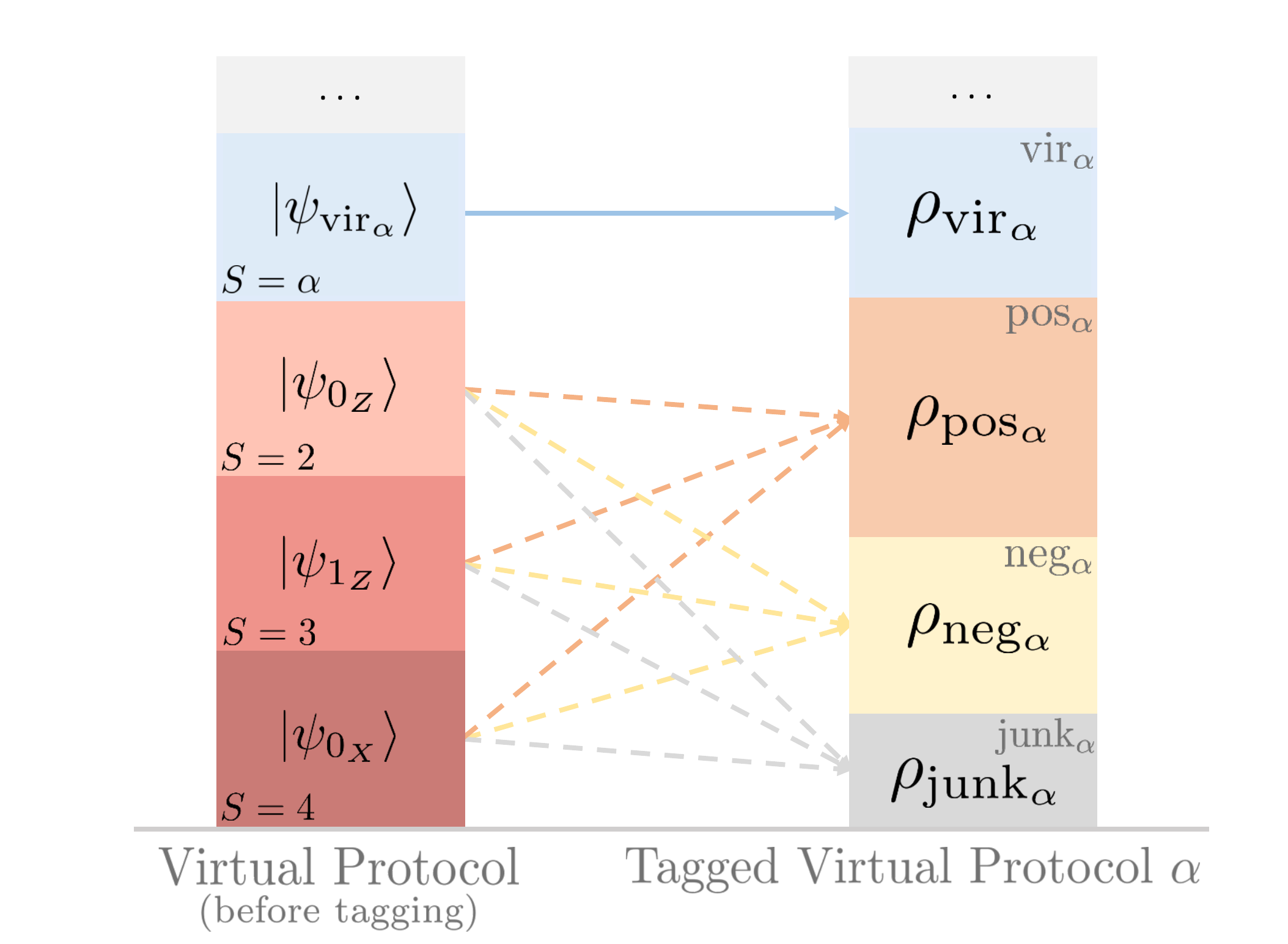}
		\caption{Relation between the virtual protocol and the Tagged Virtual Protocol $\alpha$, where $\alpha \in \{0,1\}$, for the P\&M scheme. In the virtual protocol, events for which $S \in \{2,3,4\}$ are probabilistically assigned a tag of $t_\alpha \in \{\rm pos_{\alpha},neg_{\alpha},junk_{\alpha}\}$ (dashed arrows), in such a way that the average state with a tag of $t_\alpha$ is $\rho_{t_\alpha}$. Events for which $S=\alpha$ are deterministically assigned a tag of $t_\alpha = {\rm vir_\alpha}$ (solid arrow). If one considers only the tags of $t_\alpha$ that Alice has assigned, the virtual protocol becomes equivalent to the Tagged Virtual Protocol $\alpha$. The ellipses at the top of the diagram represent events which are identical in both scenarios, but which are not relevant for the analysis.} \label{fig:pm}
	\end{figure}

After these tag assignments, an emission with a tag of $t_{\alpha}$ is equivalent to an emission of $\rho_{t_\alpha}$. Thus, if Alice disregards the outcome of her measurement of system $S$, and considers only the tags of $t_{\alpha}$ that she assigns, the virtual protocol becomes equivalent to a scenario in which Alice actually emits $\rho_{t_\alpha}$ with probability $p_{t_\alpha}$, and then trivially assigns her emission a tag of $t_{\alpha}$. This scenario, which we denote as the the Tagged Virtual Protocol $\alpha$ and depict on the right side of \cref{fig:pm}, is identical to the TVP defined in \cref{sec:general} and shown on the left side of \cref{fig:diagram}.

Let $N^{1_X}_{t_0}$ ($N^{0_X}_{t_1}$) be the number of detected events with a tag of $t_0$ ($t_1$) in which Bob obtained measurement result $1_X$ ($0_X$). \Cref{eq:Nvir4} of \cref{sec:general} implies that, in the Tagged Virtual Protocol 0, it holds that, except with probability $\varepsilon/2$,
\begin{equation}
\label{eq:N_vir0_1x}
N_{\rm vir0}^{1_X} \leq g_U \left(N_{\rm pos0}^{1_X} - g_L\Big(N_{\rm neg0}^{1_X},\tilde{p}_{{\rm pos0} | \rho_{\rm neg0}},\varepsilon/4 \Big), \tilde{p}_{{\rm vir0} | \rho_{\rm vir0}}, \varepsilon/4  \right),
\end{equation}
and in the Tagged Virtual Protocol 1, it holds that, except with probability $\varepsilon/2$,
\begin{equation}
\label{eq:N_vir1_0x}
N_{\rm vir1}^{0_X} \leq g_U \left(N_{\rm pos1}^{0_X} - g_L\Big(N_{\rm neg1}^{0_X},\tilde{p}_{{\rm pos1} | \rho_{\rm neg1}},\varepsilon/4 \Big), \tilde{p}_{{\rm vir1} | \rho_{\rm vir1}}, \varepsilon/4  \right),
\end{equation}
where, for $\alpha \in \{0,1\}$, $\tilde{p}_{{\rm vir\alpha} | \rho_{\rm vir\alpha}} = p_{\rm vir\alpha}/(p_{\rm vir\alpha} + p_{\rm pos\alpha}/c_{\rm pos\alpha})$ and $\tilde{p}_{{\rm pos\alpha} | \rho_{\rm neg\alpha}} = 1 - p_{\rm neg\alpha}/(p_{\rm neg\alpha} + p_{\rm pos\alpha}c_{\rm neg\alpha}/c_{\rm pos\alpha})$. Moreover, since the virtual protocol is equivalent to the Tagged Virtual Protocol 0 (1), in terms of the assigned tags of $t_0$ ($t_1$), \cref{eq:N_vir0_1x} (\cref{eq:N_vir1_0x}) must also hold for the virtual protocol. Thus, combining \cref{eq:N_vir0_1x,eq:N_vir1_0x}, we have that, in the virtual protocol, the number of phase errors $N_{\rm ph} := N_{\rm vir0}^{1_X} + N_{\rm vir1}^{0_X}$ satisfies
\begin{equation}
\label{eq:Nph_pm}
\begin{aligned}
	N_{\rm ph} \leq & \ g_U \left(N_{\rm pos0}^{1_X} - g_L\Big(N_{\rm neg0}^{1_X},\tilde{p}_{{\rm pos0} | \rho_{\rm neg0}},\varepsilon/4 \Big), \tilde{p}_{{\rm vir0} | \rho_{\rm vir0}}, \varepsilon/4  \right)  \\
 	+& \ g_U \left(N_{\rm pos1}^{0_X} - g_L\Big(N_{\rm neg1}^{0_X},\tilde{p}_{{\rm pos1} | \rho_{\rm neg1}},\varepsilon/4 \Big), \tilde{p}_{{\rm vir1} | \rho_{\rm vir1}}, \varepsilon/4  \right),
\end{aligned}
\end{equation}
except with probability $\varepsilon$.
	
In order to use \cref{eq:Nph_pm} to prove the security, the quantities $N_{t_0}^{1_X}$ and $N_{t_1}^{0_X}$, for $\alpha \in \{0,1\}$ and $t_\alpha \in \{{\rm pos_{\alpha}}, {\rm neg_{\alpha}}\}$, must be observables in an actual implementation of the protocol. Thus, the probabilistic tag assignments  defined in \cref{eq:tag_probability} must happen in the actual protocol too. However, note the following: (1) the tag assigned to a particular emission must be independent of Bob's measurement result, since the tag assignment does not change the emitted quantum state; and (2) the assignment of tag $t_{\alpha}$ is only relevant for the analysis if Bob happens to obtain a measurement outcome of $(\alpha \oplus 1)_X$ in that round. This implies that it is only necessary for Alice to probabilistically assign a tag of $t_0$ ($t_1$) to the events in which she sent $\ket{\psi_j}_a$, $j \in \{ 0_Z, 1_Z, 0_X\}$, and Bob obtained measurement result $1_X$ ($0_X$). For a full description of the protocol, including the tagging step, see \cref{app:protocol_description_pm}.

\section{Measurement-device-independent protocol}
\label{sec:mdi}

In this section, we apply our analysis to the LT MDI QKD protocol.  For each round, Alice (Bob) selects the state $\ket{\psi_j}_a$ ($\ket{\psi_s'}_b$) with probability $p_j$ ($p_s'$), where $j$ ($s$) $\in \{ 0, 1, \tau\}$, and sends it to an untrusted middle node Charlie. As in the P\&M case, the only assumption required to apply our analysis is that all states emitted by Alice (Bob) are in the same qubit space. For simplicity, in this discussion we assume that all states lie in the $XZ$ plane of the Bloch sphere; in Appendix D, we show how to treat the case in which they do not. Emissions for which $j \in \{0,1\}$ ($s \in \{0,1\}$) are considered to belong to the $Z$ basis, and for simplicity their selection probability is assumed to be equal, i.e.\ $p_0 = p_1 =  p_Z/2$ ($p_0' = p_1' =  p_Z'/2$). We denote Alice and Bob's joint state by $\ket{\psi_{j,s}}_{ab}\equiv\ket{\psi_{j}}_a\otimes\ket{\psi_{s}'}_b$, and its associated probability by $p_{j,s} \equiv p_j p_s'$. 

Alice and Bob expect Charlie to perform a Bell state measurement on each incoming joint pulse, and announce the result. In most MDI protocols, including the original MDI QKD proposal \cite{lo2012measurement}, Charlie may obtain a projection to one of two Bell states. However, for simplicity, for now we assume that Charlie attempts to obtain a projection to only one of the four Bell states, and that if he is successful (unsuccessful), he reports the round as "detected" ("undetected"). At the end of the section, we show how to generalise the analysis to the case in which Charlie may report a projection to two or more different Bell states.
Also, note that Charlie is untrusted, and may even be fully controlled by Eve. Thus, in what follows, we directly assume that it is Eve who performs the measurements and announces the results. Importantly, Eve is not limited to measuring each round independently: if she performs a coherent attack, her full set of announcements may depend on an arbitrary general measurement acting jointly on the photonic systems of all the rounds in the protocol. 

\begin{figure}[ht]
	\centering	\includegraphics[width=0.60\textwidth]{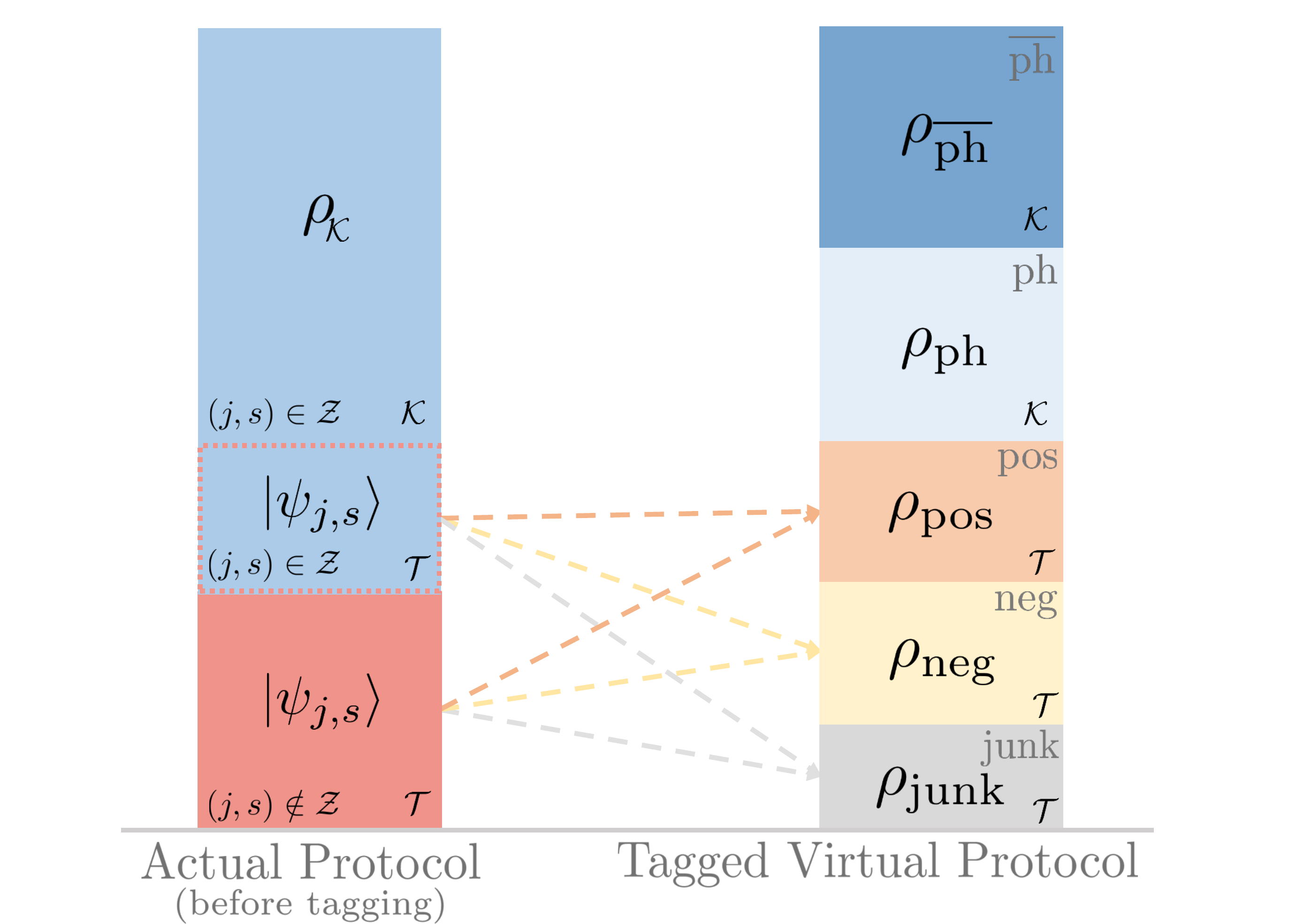}
	\caption{Relationship between the actual protocol and the tagged virtual protocol in the MDI scenario. In the actual protocol, shown on the left, emissions such that $(j,s) \in \mathcal{Z}$ are probabilistically assigned to either $\mathcal{K}$ or $\mathcal{T}$, while emissions such that $(j,s) \notin \mathcal{Z}$ are always assigned to $\mathcal{T}$. In both the actual and virtual protocols, events in $\mathcal{T}$ are probabilistically assigned a tag of $t \in \{\rm pos, neg, junk\}$, in such a way that the average state with a tag of $t$ is $\rho_t$. The dashed arrows represent this tagging process. In the virtual protocol, $\mathcal{K}$-emissions are substituted by emissions of $\rho_{\rm ph}$ and $\rho_{\rm \overline{ph}}$, and are assigned tags of "ph" and "$\overline{\rm ph}$", respectively. If Alice and Bob consider only the tags that they have assigned, the virtual protocol becomes equivalent to the \textit{tagged} virtual protocol, shown on the right.} \label{fig:tableMDI}
\end{figure}

After Eve's announcements, Alice and Bob reveal, for each round, whether or not they used the $Z$ basis, thus learning whether or not $(j,s)\in\mathcal{Z} :=\set{(0,0),(0,1),(1,0),(1,1)}$. The rounds for which $(j,s)\notin\mathcal{Z}$ are automatically considered to belong to the set of test emissions, which we denote as $\mathcal{T}$. The rounds for which $(j,s)\in\mathcal{Z}$ receive a special treatment: with probability $p_{\mathcal{K}|\mathcal{Z}}$ they are considered key emissions, and with probability $p_{\mathcal{T}|\mathcal{Z}}$ they are considered test emissions, where $\mathcal{K}$ is the set of key emissions, and $p_{\mathcal{K}|\mathcal{Z}} + p_{\mathcal{T}|\mathcal{Z}} = 1$. This is needed because we want to use data from some $\mathcal{Z}$-rounds to estimate the phase-error rate. The resulting scenario is shown on the left-hand side of \cref{fig:tableMDI}. For all rounds in $\mathcal{T}$, Alice and Bob reveal their choice of $(j,s)$.

Alice (Bob) defines her (his) sifted key as her (his) choices of $j$ ($s$) in  the detected rounds in $\mathcal{K}$. The objective of the analysis is to use the detection statistics of the $\mathcal{T}$-rounds to estimate the number of phase errors in their sifted keys. This quantity is defined as the number of errors that Alice and Bob would have obtained if they had run a virtual scenario in which they replaced the $\mathcal{K}$-emissions by the generation of the virtual state $\ket{\Psi_{\mathcal{K}}}=\frac{1}{2}\sum_{j,s=0,1}\ket{j_Z,s_Z}_{AB}\ket{\psi_{j,s}}_{ab}$, followed by an $X$-basis measurement on their local ancillas $A$ and $B$. Let $\Pi^{\rm ph}_{AB}$ be the projector onto the phase-error subspace in $AB$. Note that the definition of a phase error depends on the particular Bell state onto which Charlie is supposed to project the incoming pulses.
The average state of a key emission may be written as
\begin{equation}\label{eq:rho_key}
	\rho_{\rm \mathcal{K}} =\frac{1}{4} \sum_{j,s =0,1}  \dyad{\psi_{j,s}}{\psi_{j,s}}_{ab} = p_{\rm ph|\mathcal{K}} \rho_{\rm ph} + p_{\rm \overline{ph}|\mathcal{K}} \rho_{\rm \overline{ph}}.
\end{equation}
where $\rho_{\rm ph}$ and $\rho_{\rm \overline{ph}}$ are quantum states such that $p_{\rm  ph|\mathcal{K}}\rho_{\rm ph}=\Tr_{AB}[\Pi^{\rm ph}_{AB} \dyad*{\Psi_{\mathcal{K}}}]$ and  $p_{\rm \overline{ph}|\mathcal{K}}\rho_{\rm \overline{ph}}=\Tr_{AB}[(\mathds{I} - \Pi^{\rm ph}_{AB}) \dyad*{\Psi_{\mathcal{K}}}]$. 
Thus, the virtual protocol may be regarded as the following scenario: the users jointly select $\mathcal{K}$ or $\mathcal{T}$ with probabilities $p_{\mathcal{K}} = p_{\mathcal{Z}} p_{\mathcal{K}\vert \mathcal{Z}}$ and $p_{\mathcal{T}}= 1-p_{\mathcal{K}}$, respectively, and 

\begin{itemize}
	\item If they select $\mathcal{K}$, they emit $\rho_{\rm ph}$ and $\rho_{\rm \overline{ph}}$ with probabilities $p_{\rm ph|\mathcal{K}}$ and $p_{\rm \overline{ph}|\mathcal{K}}$, respectively.
	\item If they select $\mathcal{T}$, they emit $\ket{\psi_{j,s}}_{ab}$ with probability $p_{j,s\vert \mathcal{T}} = p_{j,s}p_{\mathcal{T}\vert j,s}/p_{\mathcal{T}}$, where $p_{\mathcal{T}\vert j,s} = p_{\mathcal{T}|\mathcal{Z}}$ if $(j,s)\in\mathcal{Z}$ and  $p_{\mathcal{T}\vert j,s} = 1$ if $(j,s)\notin\mathcal{Z}$.
\end{itemize}

The number of phase errors, $N_{\rm ph}$, is defined as the number of detected emissions of $\rho_{\rm ph}$ that Alice and Bob would have observed if they had run this virtual protocol. To estimate this quantity, we use again the random sampling analysis of \cref{sec:general}. To apply this analysis, however, we need to first show that $\rho_{\rm ph}$ can be written in the form of \cref{eq:rho_vir}, i.e.,
\begin{equation}
	\label{eq:rho_ph}
	\rho_{\rm ph} = c_{\rm pos} \rho_{\rm pos} -  c_{\rm neg} \rho_{\rm neg},
\end{equation}
and then add a tagging step to the protocol, so that it becomes equivalent to a scenario in which the states $\rho_{\rm pos}$ and $\rho_{\rm neg}$ are actually emitted. In \cref{app:coef}, we show that $\rho_{\rm ph}$ can be expressed as an operator-form linear function of the actual states, that is
\begin{equation}
\label{eq:rho_ph_js}
    \rho_{\rm ph} = \sum_{j,s}c_{j,s}\rho_{j,s},
\end{equation}
where $\rho_{j,s} \equiv \dyad{\psi_{j,s}}{\psi_{j,s}}_{ab}$ and the coefficients $c_{j,s}$ are real and can be negative. Thus, if we denote by $\mathcal{S}_{\text{pos}}$ ($\mathcal{S}_{\text{neg}}$) the set of pairs $(j,s)$ such that $c_{j,s}$ is positive (negative), and, for $t \in \{\rm pos,neg\}$, we set
\begin{align}
	c_t &= \sum_{j,s \in \mathcal{S}_t} \abs{c_{j,s}}, \\
	\rho_{t} &= \sum_{j,s \in \mathcal{S}_t}  p_{j,s|t}  \rho_{j,s},
\end{align}
where
\begin{equation}
	\label{eq:pjst}
	p_{j,s|t} :=
	\begin{cases}
		\abs{c_{j,s}}/c_{t} & \text{if $(j,s)\in\mathcal{S}_t$,} \\
		0 & \text{otherwise,}
	\end{cases}
\end{equation}
we obtain \cref{eq:rho_ph}.

In the tagging step, Alice and Bob need to probabilistically assign tags of "pos" and "neg" to their emissions in $\mathcal{T}$, in such a way that the average state with a tag of $t$ is $\rho_t$. To achieve this, in the actual protocol, Alice and Bob must assign a tag of $t \in \{\rm pos,neg\}$ to each emission of $\ket{\psi_{j,s}}_{ab}$ in $\mathcal{T}$ with probability
\begin{equation}
	\label{eq:tag_probability_MDI}
	p_{t|j,s,\mathcal{T}} = \frac{p_{t|\mathcal{T}} p_{j,s|t}}{p_{j,s|\mathcal{T}}},
\end{equation}
where $p_{j,s|t}$ is given by \cref{eq:pjst}, and $p_{t|\mathcal{T}}$ is the probability that a round in $\mathcal{T}$ is assigned a tag of $t$. Note that the assignment probabilities $p_{t|j,s,\mathcal{T}}$ become fixed once one chooses $p_{t|\mathcal{T}}$. From \cref{eq:tag_probability_MDI}, it follows that the value of $p_{t|\mathcal{T}}$ must be such that $p_{t|\mathcal{T}} \leq p_{j,s|\mathcal{T}}/p_{j,s|t}$, $\forall (j,s) \in \mathcal{S}_t$. Hence, its maximum possible value is
\begin{equation}\label{eq:p_tjsT}
	p_{t|\mathcal{T}} = \min_{j,s \in \mathcal{S}_t} \frac{p_{j,s|\mathcal{T}}}{p_{j,s|t}},
\end{equation}
and we assume that Alice and Bob choose this value, in order to waste as few $\mathcal{T}$-rounds as possible and thus obtain a tight estimate of the phase-error rate. Finally, Alice and Bob assign the tag $\textrm{"junk"}$ to all the remaining rounds in $\mathcal{T}$ that have not been tagged as $\textrm{"pos"}$ or $\textrm{"neg"}$.

Since $\mathcal{T}$-emissions are identical in the actual and virtual protocols, the previous tag assignments can be regarded as taking place in both protocols. Besides, let us further assume that, in the virtual protocol, Alice and Bob assign trivial tags of "ph" and "$\overline{\rm ph}$" to each emission of $\rho_{\rm ph}$ and $\rho_{\overline{\rm ph}}$, respectively. Then, if Alice and Bob disregard their choice of state, and consider only the tags that they have assigned, the resulting \textit{tagged} virtual protocol becomes equivalent to the scenario depicted in the right-hand side of \cref{fig:tableMDI}, in which Alice and Bob emit $\rho_{t}$, $t\in\set{\rm{ph},\overline{\rm{ph}},\textrm{pos},\textrm{neg},\textrm{junk}}$, with probability $p_t$; where $p_t=p_{\mathcal{K}}p_{t|\mathcal{K}}$ for $t\in\set{\rm{ph},\overline{\rm{ph}}}$, and  $p_t=p_{\mathcal{T}}p_{t|\mathcal{T}}$ for $t\in\set{\textrm{pos},\textrm{neg},\textrm{junk}}$. This scenario is identical to the starting point of the random sampling analysis in \cref{sec:general}, the TVP shown on
the left side of \cref{fig:diagram}. The only differences are that here we have denoted the virtual state of interest as $\rho_{\rm ph}$, not $\rho_{\rm vir}$; and that we have some extra emissions of $\rho_{\rm \overline{ph}}$ and $\rho_{\rm junk}$, which we simply ignore in the analysis. Using \cref{eq:Nvir4}, we have that, except with probability $\varepsilon$, the number of phase errors $N_{\rm ph}$ satisfies
\begin{equation}
	\label{eq:Nvir4MDI}
	N_{\rm ph} \leq g_U \left(N_{\rm pos} - g_L\Big(N_{\rm neg},\tilde{p}_{{\rm pos} | \rho_{\rm neg}},\varepsilon/2 \Big), \tilde{p}_{{\rm ph} | \rho_{\rm ph}}, \varepsilon/2  \right),
\end{equation}
where $N_{t}$ is the number of detected events with a tag of $t$, $\tilde{p}_{{\rm ph} | \rho_{\rm ph}} = p_{\textrm{ph}}/(p_{\rm ph} + p_{\rm pos}/c_{\rm pos})$ and $\tilde{p}_{{\rm pos} | \rho_{\rm neg}} = 1 - p_{\textrm{neg}}/(p_{\rm neg} + p_{\rm pos}c_{\rm neg}/c_{\rm pos})$.

In the analysis above, we have assumed that Alice and Bob reveal their choice of basis for all rounds, and then probabilistically assign all events such that $(j,s)\in\mathcal{Z}$ to either $\mathcal{T}$ or $\mathcal{K}$ with probabilities $p_{\mathcal{T}|\mathcal{Z}}$ and $p_{\mathcal{K}|\mathcal{Z}}$. However, note the following: (1) the probability to assign a particular emission to $\mathcal{T}$ or $\mathcal{K}$ must be independent of whether or not it is detected, since Eve has no information about this assignment when she makes her announcements; and (2) the set assigned to the undetected rounds is irrelevant, since their data is not used at any point in the analysis. This implies that it is only necessary for Alice and Bob to reveal their choice of basis in the detected rounds, and then assign each detected event such that $(j,s)\in\mathcal{Z}$ to either $\mathcal{T}_{\rm d}$ or $\mathcal{K}_{\rm d}$ with probabilities $p_{\mathcal{T}|\mathcal{Z}}$ and $p_{\mathcal{K}|\mathcal{Z}}$, respectively, where $\mathcal{T}_{\rm d}$ ($\mathcal{K}_{\rm d}$) is the set of detected test (key) rounds. By a similar argument, we conclude that Alice and Bob only need to reveal their choice of $(j,s)$ for the emissions in $\mathcal{T}_{\rm d}$, and then assign each of them a tag of $t \in \{\rm pos,neg\}$ with probability $p_{t|j,s,\mathcal{T}}$. For a full description of the protocol, including these assignments, see \cref{app:protocol_description_mdi}.

\subsection*{Case in which Charlie reports several projections}

The analysis above can be easily generalised to the case in which Charlie may report a projection to two or more Bell states. Essentially, the procedure is simply repeated separately for each successful projection announcement $\Omega$. Note that, because the definition of a phase error depends on $\Omega$, so does the operator associated with a phase error, which we now denote as $\rho_{\rm ph_{\Omega}}$. By repeating the procedure in Eqs.~(\ref{eq:rho_ph}) to (\ref{eq:pjst}), we define the operators $\rho_{\rm pos_{\Omega}}$ and $\rho_{\rm neg_{\Omega}}$, and the coefficients $c_{\rm pos_{\Omega}}$ and $c_{\rm neg_{\Omega}}$, for each $\Omega$. Then, we imagine that, for all $\Omega$, Alice and Bob assign a tag $t_{\Omega} \in \{\rm pos_{\Omega}, neg_{\Omega}\}$ to each emission in $\mathcal{T}$ with probability $p_{t_\Omega|j,s,\mathcal{T}}$, defined similarly to \cref{eq:p_tjsT}, in such a way that the average state with a tag of $t_{\Omega}$ is $\rho_{t_{\Omega}}$. In the virtual protocol, we also imagine that Alice and Bob assign a tag $t_{\Omega} = {\rm ph_{\Omega}}$ to each emission of $\rho_{\rm ph_\Omega}$. Then, if Alice and Bob look only at the assigned tag of $t_{\Omega}$, the scenario becomes equivalent to the "Tagged Virtual Protocol $\Omega$", in which Alice and Bob emit $\rho_{t_{\Omega}}$ with probability $p_{t_{\Omega}}$. Let $N_{t_\Omega}$ be the number of events with a tag of $t_{\Omega}$ in which Charlie announced $\Omega$. Applying the results of \cref{sec:general} to the "Tagged Virtual Protocol $\Omega$", we have that, except with probability $\varepsilon_{\Omega}$,
\begin{equation}
	\label{eq:Nvir4MDIOmega}
	N_{\rm ph_\Omega} \leq g_U \left(N_{\rm pos_\Omega} - g_L\Big(N_{\rm neg_\Omega},\tilde{p}_{{\rm pos_\Omega} | \rho_{\rm neg_\Omega}},\varepsilon_{\Omega}/2 \Big), \tilde{p}_{{\rm ph_\Omega} | \rho_{\rm ph_\Omega}} , \varepsilon_{\Omega}/2 \right) := N_{\rm ph_\Omega}^{\rm U},
\end{equation}
and because of the equivalence between the "Tagged Virtual Protocol $\Omega$" and the virtual protocol, \cref{eq:Nvir4MDIOmega} must also hold for the latter, for all $\Omega$. Thus, the total number of phase errors is upper bounded by
\begin{equation}
    N_{\rm ph} \leq \sum_{\Omega}  N_{\rm ph_\Omega}^{\rm U},
\end{equation}
except with probability $\varepsilon = \sum_{\Omega} \varepsilon_{\Omega}$. By a similar argument as in the main analysis above, we deduce that, in the actual protocol: (1) Alice and Bob only need to reveal their choice of basis in the detected rounds, and then assign each detected event such that $(j,s)\in\mathcal{Z}$ to either $\mathcal{T}_{\rm d}$ or $\mathcal{K}_{\rm d}$ with probabilities $p_{\mathcal{T}|\mathcal{Z}}$ and $p_{\mathcal{K}|\mathcal{Z}}$, respectively, where $\mathcal{T}_{\rm d}$ ($\mathcal{K}_{\rm d}$) is the set of detected test (key) rounds; and (2) Alice and Bob only need to reveal their choice of $(j,s)$ for the emissions in $\mathcal{T}_{\rm d}$, and then assign each of them a tag of $t_{\Omega} \in \{\rm pos_{\Omega},neg_{\Omega}\}$ with probability $p_{t_\Omega|j,s,\mathcal{T}}$, where $\Omega$ is Charlie's announcement on that round.

\section{Secret-key rate and security parameter}
\label{sec:security_bounds}

In \cref{sec:pam,sec:mdi}, we have shown how to obtain an upper bound $N_{\rm ph}^{\rm U}$ on the number of phase errors $N_{\rm ph}$ such that 
\begin{equation}
\label{eq:NphU_skr}
    \Pr \big[N_{\rm ph}^{\rm U} >  N_{\rm ph}\big] \leq \varepsilon.
\end{equation}
After calculating this bound, Alice and Bob perform error correction, error verification, and privacy amplification. They obtain a secret key of length
\begin{equation}
\label{eq:secretkeyrate}
    K = N_{\rm s} (1- h(N_{\rm ph}/N_{\rm s})) - \lambda_{\rm EC} - \log_2 \frac{1}{\epsilon_{\rm c}} - \log_2 \frac{1}{\xi},
\end{equation}
where $N_{\rm s}$ is the length of the sifted key, $\lambda_{\rm EC}$ is the number of bits revealed in the error correction step, and $\epsilon_{\rm c}$ is the probability that Alice and Bob's keys will not be identical after the error verification step. It is known \cite{hayashi2012concise,maeda2019repeaterless} that, if the number of phase errors is bounded as in \cref{eq:NphU_skr} and the secret-key length is set as in \cref{eq:secretkeyrate}, then the protocol is $\epsilon_{\rm s}$-secret, with $\epsilon_s = \sqrt{2} \sqrt{\varepsilon+\xi}$. Since the protocol is also $\epsilon_{\rm c}$-correct, then it is $\epsilon_{\rm sec}$-secure, with $\epsilon_{\rm sec} = \epsilon_{\rm c} + \epsilon_{\rm s}$.

\section{Numerical results}
\label{sec:numerical_results}

In this section, we simulate the secret key obtainable for both the P\&M and MDI LT protocols, using the analysis introduced in the previous sections. As usual, we assume the nominal scenario in which no eavesdropper is present. Moreover, we assume that the users' sources emit three different imperfectly-encoded single-photon states in the form
\begin{align}\label{eq:transmitted_states}
    \ket{\psi_j}=\cos(\theta_j)\ket{0_Z}+\sin(\theta_j)\ket{1_Z},
\end{align}
where $\{\ket{0_Z}, \ket{1_Z}\}$ forms a qubit basis, and $\theta_j \in [0,2\pi)$ is the encoded phase. For the P\&M scheme, we assume that Alice's states satisfy $\theta_{0Z} =0$, $\theta_{1Z}=\kappa \pi/2$, and $\theta_{0X}=\kappa \pi/4$,  where $\kappa=1+\delta/\pi$ and $\delta$ quantifies the magnitude of the SPFs. For the MDI setup, we assume that Alice's and Bob's states satisfy $\theta_{0} = \theta'_{0} = 0$,  $\theta_{1} = \theta'_{1} = \kappa\pi/2$, $\theta_{\tau} = \kappa\pi/4$ and $\theta'_{\tau} = -\kappa\pi/4$, where $\theta_j$ ($\theta'_s$) denotes the angle of Alice's (Bob's) state when she (he) emits state $j$ ($s$).

\begin{figure}[ht]
		\centering
         \subfloat[P\&M Protocol]{\includegraphics[width=0.50\textwidth]{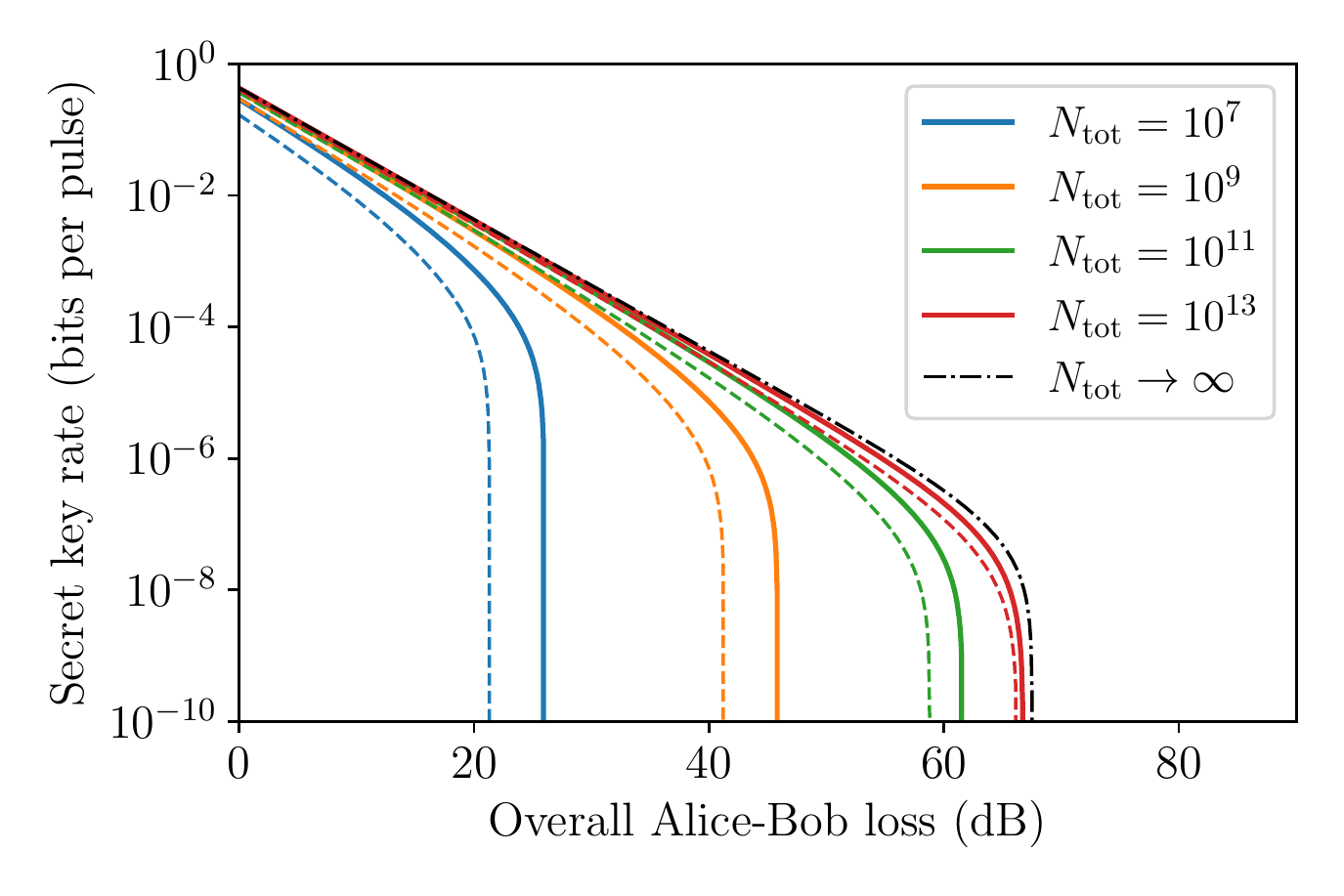}}
         \subfloat[MDI Protocol]{\includegraphics[width=0.50\textwidth]{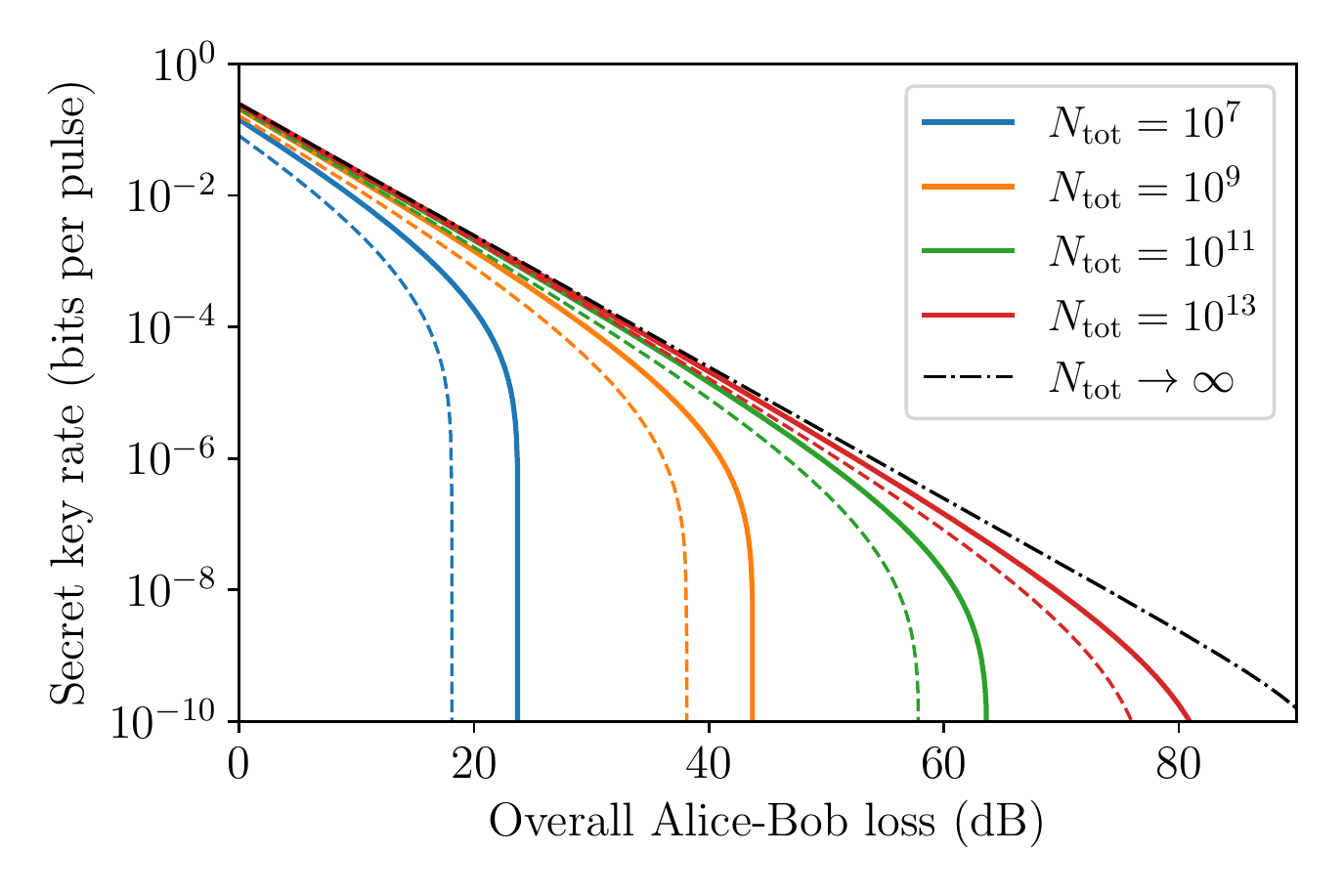}}
		\caption{Secret-key rate obtainable using our analysis based on random sampling theory (solid lines), for the P\&M (a) and MDI (b) LT protocols, as a function of the overall channel loss and for different values of the block size $N_{\rm tot}$. For comparison, we include the secret-key rate obtainable using an alternative analysis based on Azuma's inequality (dashed lines), similar to that of Ref.~\cite{mizutani2}. For both LT protocols, our analysis clearly outperforms the alternative analysis based on Azuma's inequality. 
		} \label{fig:results_azuma}
	\end{figure}

To simulate the data that would be obtained in an experiment, we use the channel model in Ref.~\cite{pereira} for the P\&M protocol, and the channel model in \cref{app:channelModelMDI} for the MDI protocol. For simplicity, in the latter we assume that Charlie only announces a detection if he obtains a projection to the Bell state $\Psi^-$. The experimental parameters considered are: SPF's parameter $\delta = 0.126$, error correction inefficiency $f = 1.16$, dark count probability of the detectors $p_d = 10^{-8}$ and fiber loss coefficient $\alpha = 0.2$ dB/km. Moreover, we select the correctness and secrecy parameters to be $\epsilon_c = 10^{-8}$ and $\epsilon_{\rm s} = 10^{-8}$, respectively, and for simplicity we set $\xi = \varepsilon$ in \cref{eq:secretkeyrate}, which means that $\varepsilon = \epsilon_{\rm s}^2/4$. In our simulations, we optimise over Alice and Bob's basis selection probabilities, and in the MDI protocol, we also optimise over the value of $p_{\mathcal{T}|\mathcal{Z}}$. Also, we consider different values of the block size $N_{\rm tot}$, which represents the total number of rounds in the protocol. Finally, we assume an error-correction leakage of $\lambda_{\rm EC} = f h(e_Z)$ bits, where $e_Z$ is the bit-error rate of the sifted key. The results for the P\&M and the MDI LT protocols are shown in \cref{fig:results_azuma}(a) and \cref{fig:results_azuma}(b), respectively.

For completeness, we compare our results with those of an alternative analysis based on the application of Azuma's inequality. This alternative analysis, presented in \cref{app:analysis_dependent_inequalities}, is essentially a simplified version of the security proof in Ref.~\cite{mizutani2}, which considers the emission of weak coherent pulses rather than single photons. The results in \cref{fig:results_azuma} show that our analysis based on random sampling offers significantly higher performances for both the P\&M and MDI LT protocols. The difference in performance is larger for lower values of $N_{\rm tot}$, while as $N_{\rm tot}$ increases, the two analyses slowly converge. In the case $N_{\rm tot} \rightarrow \infty$, both analyses provide a perfect estimation of the phase-error rate, and thus offer the same secret-key rate.

We note that a novel concentration inequality for sums of dependent random variables has been recently uploaded to a preprint server by Kato \cite{kato}. This result can be regarded as an improved version of Azuma's inequality that is much tighter when the success probability of the random variables is low. In \cref{app:analysis_dependent_inequalities}, we give a statement of the result, and use it to substitute Azuma's inequality in the alternative finite-key analysis of the LT protocol. However, it must be said that, when applied to QKD protocols, Kato's inequality requires an extra condition that is not needed in either our analysis based on random sampling or analyses based on Azuma's inequality. Namely, it requires users to attempt to predict the results that they expect to obtain in the experiment, before they actually run the experiment. This is an important step, since the inequality is only tight when the actual experimental data was reasonable close to their predictions \cite{curraslorenzo2019tight}.

\begin{figure}[ht]
        \subfloat[P\&M Protocol]{\includegraphics[width=0.49\textwidth]{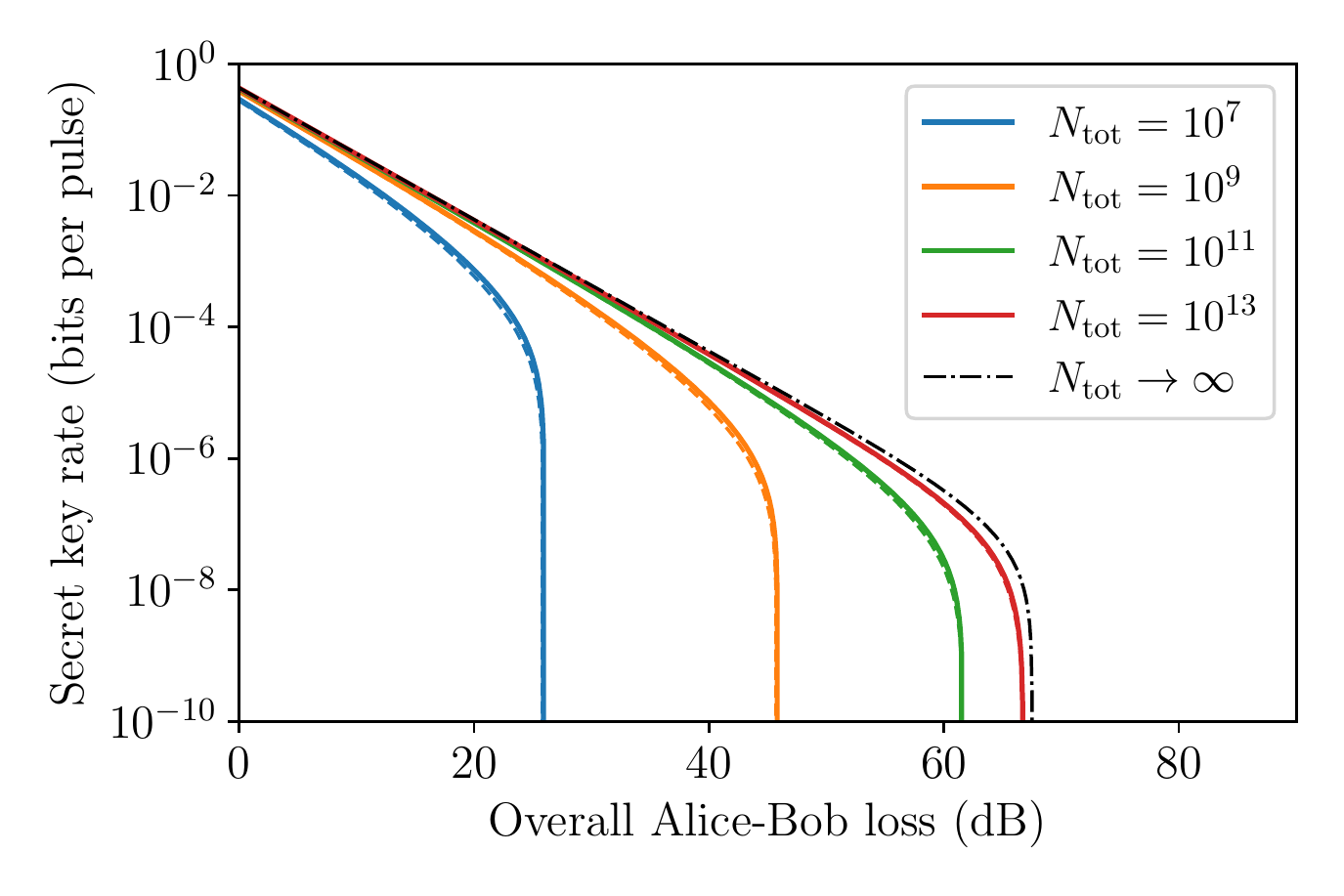}}
         \subfloat[MDI Protocol]{\includegraphics[width=0.49\textwidth]{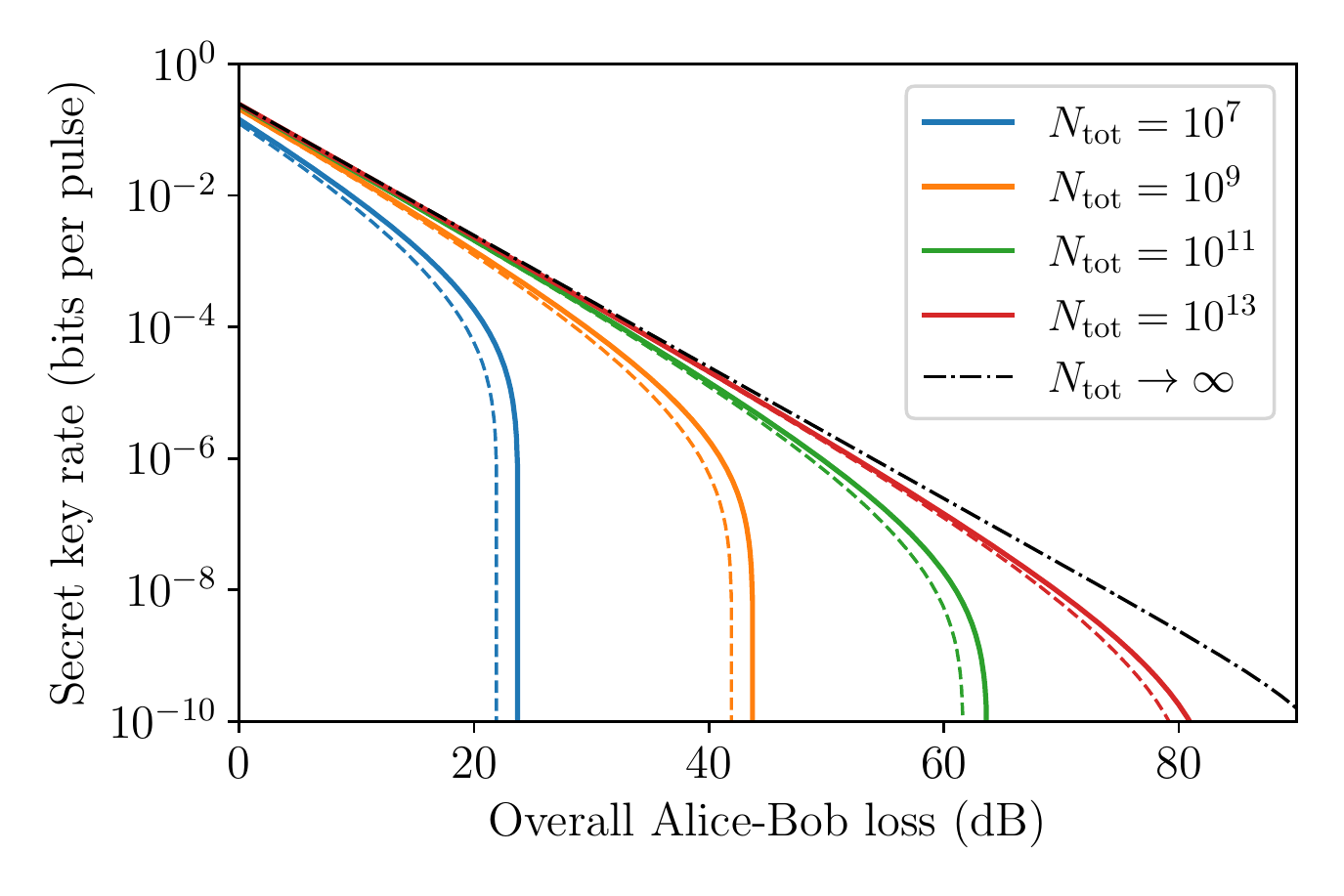}}
		\centering
		\caption{Comparison between the secret-key rate obtainable using our random sampling analysis (solid lines) and our alternative analysis based on the application of a novel concentration inequality for dependent random variables (dashed lines), for both the P\&M (a) and MDI (b) versions of the LT protocol and  different values of the block size $N_{\rm tot}$. For the P\&M protocol, the performance of the two security proofs is almost identical, while for the MDI protocol, our analysis based on random sampling provides slightly better secret-key rates.
		} \label{fig:results_kato}
	\end{figure}

In \cref{fig:results_kato}, we compare the performance of our analysis based on random sampling theory with that of our alternative analysis based on Kato's inequality. For simplicity, in the alternative analysis, we assume that the users could perfectly predict the experimental data that they obtain in the experiment, which maximises the secret-key rate obtainable. \cref{fig:results_kato}(a) shows that, in the case of the P\&M protocol, the difference between the two analyses vanishes almost completely. Conversely, \cref{fig:results_kato}(b) shows that, in the case of the MDI protocol, our analysis based on random sampling still retains an advantage, although significantly smaller than that observed in  \cref{fig:results_azuma}(b). We emphasise that, unlike the alternative analysis based on Kato's inequality, our analysis based on random sampling does not require the users to make any prediction before running the experiment.

  \section{Discussion}
  \label{sec:conclusion}
In this work, we have proved the finite-key security of the loss-tolerant (LT) QKD protocol against general attacks, for both its prepare-and-measure and measurement-device-independent versions. Our security analysis reduces the parameter estimation task to a classical random sampling problem, which can be solved using Chernoff bounds, and provides higher secret-key rates than previous results based on the application of Azuma's inequality \cite{mizutani2}. 

Although we have assumed single-photon sources, we believe that our analysis can be extended to the case in which the users employ weak coherent sources, as long as the single-photon components of the three encoded pulses satisfy the requirements of our proof, i.e.\ they are characterised and belong to the same qubit space. In that case, the users should assign tags to their emissions in such a way that \cref{eq:rho_vir} holds for their single-photon components; i.e.\ $\rho_{\rm vir}^{(1)} = c_{\rm pos} \rho_{\rm pos}^{(1)} -  c_{\rm neg} \rho_{\rm neg}^{(1)}$, where $\rho_{t}^{(1)}$ is the average quantum state of a single-photon pulse with a tag of $t$. If so, \cref{eq:Nvir4} holds, although it now has the form $N_{\rm vir}^{(1)} \leq f(N_{\rm pos}^{(1)}, N_{\rm neg}^{(1)};\varepsilon)$, where $N_{t}^{(1)}$ denotes the number of detected single-photon pulses with a tag of $t$. Note that now $N_{\rm pos}^{(1)}$ and $N_{\rm neg}^{(1)}$ are not directly observable, since the users do not know the photon number of their emissions. However, by using different laser intensities $\mu$, they are able to observe the values $\{{N_{\rm pos}^{\mu}}\}$ and $\{{N_{\rm neg}^{\mu}}\}$ for all $\mu$, where $N_t^{\mu}$ is the number of detected emissions with a tag of $t$ that originated from intensity $\mu$. Thus, they can apply the decoy-state method \cite{hwang2003quantum,lo2005decoy,ma2005practical} to obtain an upper (lower) bound on $N_{\rm pos}^{(1)}$ ($N_{\rm neg}^{(1)}$), using for example the numerical techniques introduced in Ref.~\cite{curty2014finite}.
  
Also, in our random sampling analysis, we have assumed that the three encoded states live in the same qubit space. In a future work, it would be interesting to consider if our security proof can be extended to the case in which the qubit assumption is not satisfied, due to additional imperfections such as mode dependencies \cite{pereira} or correlations between different rounds of the protocol \cite{pereira2,navarrete2021hopefully}. In that case, one can no longer derive an operator equality between the virtual and the actual states, such as e.g.\ \cref{eq:qubit_space}. Instead, one needs to find an operator dominance condition \cite{maeda2019repeaterless} between them, which is non-trivial if the side-channel states are not characterised, as assumed by Refs.~\cite{pereira,pereira2,navarrete2021hopefully}.

\section{Acknowledgements}
We thank Mohsen Razavi and Marcos Curty for valuable discussions. G.C.-L. and M.P. acknowledge support from the European Union's Horizon 2020 research and innovation programme under the Marie Sk\l{}odowska-Curie grant agreement number 675662 (project QCALL). A.N. acknowledges support from a FPU scholarship (FPU 15/03668) from the Spanish Ministry of Science, Innovation and Universities (MCIU). K.T. acknowledges support from JSPS KAKENHI Grant Numbers JP18H05237 18H05237 and JST-CREST JPMJCR 1671.

\bibliography{refs}

	\appendix
	\clearpage
	 \section{Random sampling analysis}
	 \label{app:random_sampling}
    	 Here, we prove the statements in \cref{eq:Nvir,eq:Nnegpos}, and give an expression for the functions $g_L$ and $g_U$. Let us assume that we have a population of $n$ items, where $n$ is unknown. Each item is assigned to either $\mathcal{K}_1$ with probability $p$ or to $\mathcal{K}_2$ with probability $1-p$. We know the value of $K_2 = \abs{\mathcal{K}_2}$ and we would like to obtain bounds on $K_1 = \abs{\mathcal{K}_1}$.
	 
	Let $\xi_i = 1$ if the $i$-th trial is assigned to $\mathcal{K}_2$ and $\xi_i = 0$ otherwise. We have that
	\begin{equation}
	\sum_{i=1}^{n} \xi_i = K_2.
	\end{equation}
	Clearly, $\mathbb{E}[K_2] = (1-p) n$, and therefore $n = \mathbb{E}[K_2]/(1-p)$. Using the inverse multiplicative Chernoff bound \cite{zhang2017improved,bahrani2019wavelength,curraslorenzo2019tight}, we have that
	\begin{equation}
	\begin{gathered}
	\mathbb{E}[K_2] \geq - K_2 W_0\left(-e^{\frac{\ln \varepsilon-K_2}{K_2}}\right) \\
	\mathbb{E}[K_2] \leq - K_2 W_{-1}\left(-e^{\frac{\ln \varepsilon-K_2}{K_2}}\right)
	\end{gathered}
	\end{equation}
	where $W_0$ and $W_{-1}$ are branches of the Lambert $W$ function, and each of the bounds fails with probability at most $\varepsilon$. From this and the fact that $n=K_1 + K_2$, we have that
	\begin{equation}
	\label{eq:functions_random_sampling}
	\begin{gathered}
	    K_1 = n - K_2 =\frac{\mathbb{E}[K_2]}{1-p}- K_2 \geq  \max \left( - \frac{K_2 W_0\left(-e^{\frac{\ln \varepsilon-K_2}{K_2}}\right)}{1-p} - K_2,\,0\right) 
	    =: g_L (K_2,p,\varepsilon), \\
	    K_1 = n - K_2 = \frac{\mathbb{E}[K_2]}{1-p} - K_2 \leq  - \frac{K_2 W_{-1}\left(-e^{\frac{\ln \varepsilon-K_2}{K_2}}\right)}{1-p} - K_2
	    =: g_U (K_2,p,\varepsilon),
	\end{gathered} 
	\end{equation}
    where each of the bounds fails with probability $\varepsilon$. It can be shown that $g_U$ is an increasing function of $K_2$.
    Note that \cref{eq:functions_random_sampling} is only valid for $K_2 > 0$. In the special case $K_2 =0$, we have that \cite{zhang2017improved}
    \begin{equation}
    \begin{gathered}
    g_L(0,p,\varepsilon) := 0 \\
    g_U(0,p,\varepsilon) := -\frac{\ln \varepsilon}{1-p}.
    \end{gathered}
    \end{equation}
    We note that this random sampling problem can also be solved using the method introduced in Ref.~\cite{maeda2019repeaterless}.
    
\section{Operator-form linear relationship between the virtual and actual states}
\label{app:coef}

In this Appendix, we show how to find an operator-form linear relationship between the virtual states and the actual states, see \cref{eq:qubit_space} and \cref{eq:rho_ph_js}. For simplicity, we provide first the procedure for the P\&M protocol; then, at the end of this Appendix, we show how to extend it to the MDI case. The only assumption on Alice's emitted states,  $\ket{\psi_{j}}_a$ for $j \in \{0_Z,1_Z,0_X\}$, is that they are linearly dependent, i.e.\ all three states live in the same qubit space. However, the analysis simplifies significantly if they are all in the same standard basis plane of the Bloch sphere, such as the $XZ$, $XY$ or $ZY$ plane. First, we consider this simpler case, and then provide the analysis for the general case.

\subsection{Case in which all states are in a standard basis plane}

Without loss of generality, we assume that the three states are in the $XZ$ plane of the Bloch sphere, i.e.\ they can be expressed as 
\begin{equation}
\label{eq:psi_j_XZplane}
    \ket{\psi_j}_a=\cos(\theta_j)\ket{0_Z}_a+\sin(\theta_j)\ket{1_Z}_a,
\end{equation}
where $\theta_j = (-\pi,\,\pi]$. Alice generates her sifted key from the detected emissions of $\ket{\psi_{0_Z}}_a$ and $\ket{\psi_{1_Z}}_a$. To prove the security of the sifted key, we consider an entanglement-based virtual protocol in which Alice prepares the state
\begin{equation}\label{eq:virtual_entangled_state_app}
\ket{\Psi_{Z}}_{Aa}=\frac{1}{\sqrt{2}}\left(\ket{0_Z}_A\ket{\psi_{0_Z}}_{a}+\ket{1_Z}_A\ket{\psi_{1_Z}}_{a}\right).
\end{equation}
In this virtual protocol, Alice measures her local ancilla $A$ in the complementary basis $\set{\ket{0_X}_A,\ket{1_X}_A}$, where $\ket{\alpha_X}_A=1/\sqrt{2}\left[\ket{0_Z}_A+(-1)^\alpha\ket{1_Z}_A\right]$ for $\alpha \in \{0,1\}$. If Alice obtains $\ket{\alpha_X}_A$, she effectively emits the virtual state
\begin{equation}
\label{eq:virtual_states_app}
\ket{\psi_{\rm vir_{\alpha}}}_a = \frac{1}{\sqrt{p_{\textrm{vir}_{\alpha}\vert Z}}} \big(\ket{\psi_{0_Z}}_{a} + (-1)^{\alpha} \ket{\psi_{1_Z}}_{a}\big),
\end{equation}
where $p_{\textrm{vir}_{\alpha} \vert Z} = \norm{\ket{\psi_{0_Z}}_{a}+(-1)^{\alpha} \ket{\psi_{1_Z}}_{a}}^2/4 =  (1+(-1)^{\alpha}\cos(\theta_{0_Z}-\theta_{1_Z}))/2$ is the probability that Alice obtains $\ket{\alpha_X}_A$. Since $\ket{\psi_{0_Z}}_a$ and $\ket{\psi_{1_Z}}_a$ are in the $XZ$ plane, $\ket{\psi_{\rm vir_{\alpha}}}_a$ is also in the $XZ$ plane.

Let $[S^Z_j,\, S^X_j,\, S^Y_j]$ be the Bloch vector of the state $\ket{\psi_j}_a$. We have that $S^Z_j = \cos(2 \theta_j)$,  $S^X_j = \sin(2 \theta_j)$ and $S^Y_j = 0$. Thus, in operator form, the state $\ket{\psi_j}_a$ can be expressed as
\begin{equation}
\label{eq:pauli_formula}
    \rho_j \equiv \ketbra{\psi_j}_a = \frac{1}{2} \big(\sigma_I + S^Z_j \sigma_Z + S^X_j \sigma_X\big),
\end{equation}
where $\sigma_I$ is the identity operator and $\sigma_K$, for $K \in \{Z, X, Y\}$, is a Pauli operator. It is useful to see \cref{eq:pauli_formula} as a system of linear equations, with three unknowns ($\sigma_I$, $\sigma_Z$, $\sigma_X$) and three equations (one for each $\ket{\psi_j}_a$). We can write this system in matrix form:
\begin{equation}
\label{eq:pauli_formula_matrix}
\bm{\rho} =  \bm{S} \bm{\sigma},
\end{equation}
where $\bm{\rho} = \left[\rho_{0_Z},\,\rho_{1_Z},\,\rho_{0_X}\right]^{\rm T}$, $\bm{\sigma} = \left[\sigma_I,\,\sigma_Z,\,\sigma_X\right]^{\rm T}$, and
\begin{equation}
\bm{S}:= \frac{1}{2}\begin{bmatrix}
1 & S^Z_{0_Z} & S^X_{0_Z}\\
1 & S^Z_{1_Z} & S^X_{1_Z} \\
1 & S^Z_{0_X} & S^X_{0_X}
\end{bmatrix};
\end{equation}
and express its solution as
\begin{equation}
\label{eq:pauli_formula_matrix_solution}
\bm{\sigma} = \bm{S}^{-1} \bm{\rho}.
\end{equation}
\Cref{eq:pauli_formula_matrix_solution} essentially says that the operators $\sigma_I,\,\sigma_Z,\,\sigma_X$ can be expressed as a linear combination of the actual states $\rho_j$. This implies that every state that can be expressed as a linear combination of $\sigma_I,\,\sigma_Z,\,\sigma_X$ (i.e., every state in the $XZ$ plane) can also be expressed as a linear combination of $\rho_j$. In particular, the virtual states $\ket{\psi_{\rm vir_{\alpha}}}_a$ are in the $XZ$ plane, and in operator form they can be expressed as 
\begin{equation}
\rho_{\rm vir_{\alpha}} \equiv \ketbra{\psi_{\rm vir_{\alpha}}}_a = \frac{1}{2} \big(\sigma_I + S^Z_{\rm vir_{\alpha}} \sigma_Z + S^X_{\rm vir_{\alpha}} \sigma_X\big),
\end{equation}
where $S^Z_{\rm vir_{0}} = - S^Z_{\rm vir_{1}} = \cos(\theta_{0_Z} + \theta_{1_Z})$ and $S^X_{\rm vir_{0}} = - S^X_{\rm vir_{1}} = \sin(\theta_{0_Z} + \theta_{1_Z})$. Or equivalently,
\begin{equation}
\label{eq:pauli_formula_virtual_matrix}
\rho_{\rm vir_{\alpha}} = \bm{S}_{\rm vir_{\alpha}} \bm{\sigma},
\end{equation}
where $\bm{S}_{\rm vir_{\alpha}} = \frac{1}{2}\big[1,\,S^Z_{\rm vir_{\alpha}},\,S^X_{\rm vir_{\alpha}}\big]^{\rm T}$. Combining \cref{eq:pauli_formula_matrix_solution,eq:pauli_formula_virtual_matrix}, we have that
\begin{equation}
    \rho_{\rm vir_{\alpha}} = \bm{S}_{\rm vir_{\alpha}} \bm{S}^{-1} \bm{\rho} = \bm{f}_{\alpha} \bm{\rho},
\end{equation}
where $\bm{f}_{\alpha}:= \bm{S}_{\rm vir_{\alpha}} \bm{S}^{-1}$ is a row vector. If we express $\bm{f}_{\alpha}$ as $\bm{f}_{\alpha} = [c_{0_Z \vert \alpha},\,c_{1_Z \vert \alpha},\,c_{0_X \vert \alpha}]$, we obtain \cref{eq:qubit_space}, i.e.
\begin{equation}
     \rho_{\rm vir_{\alpha}} = c_{0_Z \vert {\rm vir_\alpha}}  \rho_{0_Z}  + c_{1_Z \vert {\rm vir_\alpha}} \rho_{1_Z} + c_{0_X \vert {\rm vir_\alpha}} \rho_{0_Z},
\end{equation}
for $\alpha \in \{0,1\}$, as required.

In our numerical simulations, we assume that the three states emitted by Alice are in the $XZ$ plane, and that when written as in \cref{eq:psi_j_XZplane}, their phases satisfy $\theta_{0_Z} = 0$, $\theta_{1_Z} = \kappa\pi/2$ and $\kappa\pi/4$, for some $\kappa$. For this particular case, an analytical expression for the coefficients is given by 
\begin{equation}
\begin{aligned}
&c_{0_Z \vert {\rm vir_0}} = c_{1_Z \vert {\rm vir_0}} = 0, \nonumber \\
&c_{0_X \vert {\rm vir_0}} = 1, \nonumber \\
&c_{0_Z \vert {\rm vir_1}} = c_{1_Z \vert {\rm vir_1}} =  \csc^2(\kappa\pi/4))/2, \nonumber\\
&c_{0_X \vert {\rm vir_1}} = -\cot^2(\kappa\pi/4)).
\end{aligned}
\end{equation}

\subsection{General case}

Here, we consider the case in which the three states are not all in the same standard basis plane. Formally, we assume that for all $K \in \{Z,X,Y\}$, there is at least one $j$ such that $S_j^K \neq 0$. Therefore \cref{eq:pauli_formula} becomes
\begin{equation}
\label{eq:pauli_formula2}
    \rho_j \equiv \ketbra{\psi_j}_a = \frac{1}{2} \big(\sigma_I + S^Z_j \sigma_Z + S^X_j \sigma_X +  S^Y_j \sigma_Y\big),
\end{equation}
and now we have a system of three equations with four unknowns. We have to find a way to modify \cref{eq:pauli_formula2} such that it becomes a system with three unknowns.

For any basis $\{\ket{0_U}_a, \ket{1_U}_a\}$ of the qubit space, Alice's emitted states can be expressed as
\begin{equation}
\label{eq:states_Ubasis}
	\ket{\psi_j}_a=e^{i \gamma_j}\left(\sqrt{u_j}\ket{0_U}_a+e^{i\phi_j}\sqrt{1-u_j}\ket{1_U}_a\right).
\end{equation}
where $0 \leq u_j \leq 1$, $\gamma_j \in [0,\,2\pi)$, $\phi_j \in [0,2\pi)$. Since the end points of the three Bloch vectors associated to Alice's emitted states form a plane, there must be a basis $U$ such that $u_j$ has the same value $\forall j$. Expressed in this basis, which we denote as $\tilde{Y}$, the states are
\begin{equation}
\label{eq:states_Ubasis2}
	\ket{\psi_j}_a=e^{i \gamma_j}\left(\sqrt{\tilde{y}}\ket{0_{\tilde{Y}}}_a+e^{i\phi_j}\sqrt{1-\tilde{y}}\ket{1_{\tilde{Y}}}_a\right),
\end{equation}
for some $0\leq\tilde{y}\leq1$. Let $V$ be a unitary operator such that $V\ket{0_Y}_a = \ket{0_{\tilde{Y}}}_a$ and $V\ket{1_Y}_a = \ket{1_{\tilde{Y}}}_a$. 
$V$ can be regarded as a transformation from the set of mutually unbiased bases $Z$, $X$, $Y$ to the set of mutually unbiased bases $\tilde{Z}$, $\tilde{X}$, $\tilde{Y}$. Let us define the modified Pauli operators $\tilde{\sigma}_K = V \sigma_K V^\dagger$, for $K \in \{Z,X,Y\}$, and express the actual states in terms of these, i.e.\ 
\begin{equation}
\label{eq:modified_pauli_formula_prev}
\rho_j = \frac{1}{2} \big(\sigma_I + \tilde{S}_Z^j \tilde{\sigma}_Z + \tilde{S}_X^j \tilde{\sigma}_X + \tilde{S}_Y^j \tilde{\sigma}_Y\big).
\end{equation}
Note that the three states have the same $\tilde{Y}$ component, i.e.\ $\tilde{S}_Y^j = \tilde{S}_Y := 2\tilde{y}-1$, $\forall j$. This allows us to define the operator $\tilde{\sigma}_O = \sigma_I + \tilde{S}_Y \tilde{\sigma}_Y$, and rewrite \cref{eq:modified_pauli_formula_prev} as
\begin{equation}
\label{eq:modified_pauli_formula}
    \ketbra{\psi_j}_a = \frac{1}{2} \big(\tilde{\sigma}_O + \tilde{S}_Z^j \tilde{\sigma}_Z + \tilde{S}_X^j \tilde{\sigma}_X\big),
\end{equation}
which has a similar form as \cref{eq:pauli_formula}, i.e.\ it can be regarded as a linear system of three equations and three unknowns. If we define $\bm{\rho} = \left[\rho_{0_Z},\,\rho_{1_Z},\,\rho_{0_X}\right]^{\rm T}$, $\bm{\sigma} = \left[\tilde{\sigma}_O,\,\tilde{\sigma}_Z,\,\tilde{\sigma}_X\right]^{\rm T}$, and 
\begin{equation}
\bm{S}:= \frac{1}{2}\begin{bmatrix}
1 & \tilde{S}^Z_{0_Z} & \tilde{S}^X_{0_Z}\\
1 & \tilde{S}^Z_{1_Z} & \tilde{S}^X_{1_Z} \\
1 & \tilde{S}^Z_{0_X} & \tilde{S}^X_{0_X}
\end{bmatrix};
\end{equation}
we have that $\bm{\rho} = \bm{S} \bm{\sigma}$, and therefore,
\begin{equation}
\label{eq:modified_pauli_formula_matrix_solution}
\bm{\sigma} = \, \bm{S}^{-1} \bm{\rho}.
\end{equation}
The previous equation implies that the modified Pauli operators $\tilde{\sigma}_O,\,\tilde{\sigma}_Z,\,\tilde{\sigma}_X$ can be expressed as a linear combination of the actual states $\rho_j$. Therefore, any state that can be expressed as a linear combination of $\tilde{\sigma}_O,\,\tilde{\sigma}_Z,\,\tilde{\sigma}_X$ (i.e.\ any state whose $\tilde{Y}$-component is $\tilde{S}_Y$) can also be expressed as a linear combination of the $\rho_j$.

If we define the virtual states as in \cref{eq:virtual_states_app}, it is likely that they will not satisfy the condition that their $\tilde{Y}$-component is $\tilde{S}_Y$. However, note that to obtain \cref{eq:virtual_states_app}, we have assumed that Alice measures the ancilla $A$ of the entangled state in  \cref{eq:virtual_entangled_state_app} in the $X$ basis. In reality, Alice could have decided to measure it in any other basis that is mutually unbiased with $Z$. Equivalently, we can express this degree of freedom by assuming that Alice does measure in the $X$ basis, but defines the entangled state as 
\begin{equation} \label{eq:virtual_entangled_state_app_noXZ}
    \ket{\Psi_{Z}}_{Aa}=\frac{1}{\sqrt{2}}\left(\ket{0_Z}_A\ket{\psi_{0}}_{a}+e^{i\phi}\ket{1_Z}_A\ket{\psi_{1}}_{a}\right),
\end{equation}
for some $\phi \in [0,\,2\pi)$. Thus, the virtual states now become
\begin{equation}
\label{eq:virtual_states_app_noXZ}
\ket{\psi_{\rm vir_{\alpha}}}_a = \frac{1}{\sqrt{p_{\textrm{vir}_{\alpha}\vert Z}}} \big(\ket{\psi_{0}}_{a} + (-1)^{\alpha} e^{i\phi} \ket{\psi_{1}}_{a}\big),
\end{equation}
where $p_{\textrm{vir}_{\alpha} \vert Z} =  \norm{\ket{\psi_{0}}_{a}+(-1)^{\alpha} e^{i\phi} \ket{\psi_{1}}_{a}}^2/4$ is the probability that Alice obtains $\ket{\alpha_X}_A$. Substituting \cref{eq:states_Ubasis2} in \cref{eq:virtual_states_app_noXZ}, one can easily show that if Alice chooses $\phi = \gamma_{0_Z} - \gamma_{1_Z} +\left(\phi_{0_Z} - \phi_{1_Z} \right)/2$, then the modified Bloch vector of the virtual state $\ket{\psi_{\rm vir_{\alpha}}}_a$, $[\tilde{S}^Z_{\rm vir_{\alpha}},\,\tilde{S}^X_{\rm vir_{\alpha}},\,\tilde{S}_Y^{\rm vir_{\alpha}}]$, satisfies $\tilde{S}_Y^{\rm vir_{\alpha}} = \tilde{S}_Y$ for both $\alpha \in \{0,1\}$. Therefore
\begin{equation}
\label{eq:modified_pauli_formula_virtual}
     \rho_{\rm vir_{\alpha}} = \frac{1}{2} \big(\tilde{\sigma}_O + \tilde{S}^Z_{\rm vir_{\alpha}} \tilde{\sigma}_Z + \tilde{S}^X_{\rm vir_{\alpha}} \tilde{\sigma}_X\big),
\end{equation}
or equivalently,
\begin{equation}
\label{eq:modified_pauli_formula_virtual_matrix}
\rho_{\rm vir_{\alpha}} = \bm{S}_{\rm vir_{\alpha}} \bm{\sigma},
\end{equation}
where $\bm{S}_{\rm vir_{\alpha}} = \frac{1}{2} \big[1,\,\tilde{S}^Z_{\rm vir_{\alpha}},\,\tilde{S}^X_{\rm vir_{\alpha}}\big]^{\rm T}$. Combining \cref{eq:modified_pauli_formula_matrix_solution,eq:modified_pauli_formula_virtual_matrix}, we have that
\begin{equation}
    \rho_{\rm vir_{\alpha}} = \bm{S}_{\rm vir_{\alpha}} \bm{S}^{-1} \bm{\rho} := \bm{f}_{\alpha} \bm{\rho},
\end{equation}
where $\bm{f}_{\alpha}:= \bm{S}_{\rm vir_{\alpha}} \bm{S}^{-1}$ is a row vector. If we express $\bm{f}_{\alpha}$ as $\bm{f}_{\alpha} = [c_{0_Z \vert \alpha},\,c_{1_Z \vert \alpha},\,c_{0_X \vert \alpha}]$, we obtain \cref{eq:qubit_space}, i.e.
\begin{equation}
     \ketbra{\psi_{\rm vir_{\alpha}}}_a = c_{0_Z \vert \alpha}  \ketbra{\psi_{0_Z}}_a  + c_{1_Z \vert \alpha}  \ketbra{\psi_{1_Z}}_a + c_{0_X \vert \alpha}  \ketbra{\psi_{0_X}}_a,
\end{equation}
for $\alpha \in \{0,1\}$, as required.

\subsection{MDI protocol}

In the MDI scenario, we essentially perform the above procedure separately for Alice's and Bob's states. Let $\ket{\psi_j}_a$ ($\ket{\psi_s'}_b$), with $j$ ($s$) $\in \{ 0, 1, \tau\}$, denote Alice's (Bob's) states, and let $\rho_j \equiv \ketbra{\psi_j}$ ($\rho_s' \equiv \ketbra{\psi_s'}$) denote their operator form. Using the analysis in the previous sections, we have that
\begin{equation}
\begin{gathered}
     \rho_{\rm vir_{\alpha}} = c_{0\vert \rm vir_\alpha}  \rho_{0}  + c_{1\vert \rm vir_\alpha}  \rho_{1} + c_{0\vert \rm vir_\alpha}  \rho_{\tau}, \\
     \rho'_{\rm vir_{\beta}} = c'_{0\vert \rm vir_\beta}  \rho'_{0}  + c'_{1\vert \rm vir_\beta}  \rho'_{1} + c'_{0\vert \rm vir_\beta}  \rho'_{\tau}; \\
\end{gathered}
\end{equation}
where $\alpha,\beta \in \{0,1\}$, and $\rho_{\rm vir_{\alpha}}$ ($\rho'_{\rm vir_{\beta}}$) denotes one of Alice's (Bob's) virtual states, emitted with probability $p_{\rm vir_{\alpha} \vert \mathcal{K}}$ ($p'_{\rm vir_{\beta} \vert \mathcal{K}}$). We can define Alice and Bob's joint virtual states as
\begin{equation}
\label{eq:rho_vir_alpha_beta}
    \rho_{\rm vir_{\alpha,\beta}} = \rho_{\rm vir_{\alpha}} \otimes  \rho'_{\rm vir_{\beta}} =  \sum_{j,s} c_{j,s\vert\rm vir_ {\alpha,\beta}} \rho_{j,s},
\end{equation}
emitted with probability $p_{\rm vir_{\alpha,\beta} \vert \mathcal{K}} = p_{\rm vir_{\alpha} \vert \mathcal{K}} p'_{\rm vir_{\alpha} \vert \mathcal{K}}$; where $c_{j,s\vert \rm vir_{\alpha,\beta}} = c_{j \vert \rm vir_\alpha} c'_{s \vert \rm vir_\beta}$. Depending on Charlie's Bell state report, the definition of a phase error will change. If Charlie reports a projection to either $\Psi^{-}$ or $\Phi^{-}$, the phase-error operator is defined as
\begin{equation}
\label{eq:rho_ph_same}
    \rho_{\rm ph} = (p_{\rm vir_{0,0}} \rho_{\rm vir_{0,0}} + p_{\rm vir_{1,1}} \rho_{\rm vir_{1,1}})/p_{\rm ph},
\end{equation}
where $p_{\rm ph} = p_{\rm vir_{0,0}} +  p_{\rm vir_{1,1}}$. Conversely, if he reports a projection to either $\Psi^{+}$ or $\Phi^{+}$, the phase-error operator is defined as
\begin{equation}
\label{eq:rho_ph_diff}
    \rho_{\rm ph} = (p_{\rm vir_{0,1}} \rho_{\rm vir_{0,1}} + p_{\rm vir_{1,0}} \rho_{\rm vir_{1,0}})/p_{\rm ph},
\end{equation}
where $p_{\rm ph} = p_{\rm vir_{0,1}} +  p_{\rm vir_{1,0}}$. In any case, one can express the phase-error operator as
\begin{equation}
    \rho_{\rm ph} = \sum_{j,s} c_{j,s} \rho_{j,s},
\end{equation}
where the coefficients $c_{j,s}$ are a linear function of the coefficients $c_{j,s\vert \alpha,\beta}$, and can be obtained by substituting \cref{eq:rho_vir_alpha_beta} in either \cref{eq:rho_ph_same} or \cref{eq:rho_ph_diff}.

In our numerical simulations we assume that Alice and Bob's states are in the $XZ$ plane, and that when written as in \cref{eq:psi_j_XZplane}, their phases satisfy $\theta_{0} = \theta'_{0} = 0$,  $\theta_{1} = \theta'_{1} = \kappa\pi/2$, $\theta_{\tau} =  - \theta'_{\tau} = \kappa\pi/4$. For this particular case, we have that Alice's virtual states satisfy 
\begin{equation}
\begin{aligned}
&c_{0 \vert \rm vir_0} = c_{1 \vert \rm vir_0} = 0, \\
&c_{0 \vert \rm vir_0} = 1,\\
&c_{0 \vert \rm vir_1} = c_{1 \vert \rm vir_1} =  \csc^2(\kappa\pi/4))/2,  \\
&c_{\tau \vert \rm vir_1} = -\cot^2(\kappa\pi/4);
\end{aligned}
\end{equation}
while Bob's virtual states satisfy
\begin{equation}
\begin{aligned}
&c'_{0 \vert \rm vir_0} =  1, \\
&c'_{1 \vert \rm vir_0} = -c'_{\tau \vert \rm vir_0} = \frac{1}{1+2\cos(\kappa\pi/2)},  \\ 
&c'_{0 \vert \rm vir_1} = -\frac{\cos(\kappa\pi/2) \csc^2(\kappa\pi/4)}{2},\\
&c'_{1 \vert \rm vir_1} = \frac{\cos(\kappa/2) \csc(\kappa\pi/4)\csc(3\kappa\pi/4)}{2}, \\
&c'_{\tau \vert \rm vir_1} = \frac{\cot^2(\kappa\pi/4)}{1+2\cos(\kappa\pi/2)}.
\end{aligned}
\end{equation}

    \section{Description of the P\&M protocol}
    \label{app:protocol_description_pm}

	\begin{enumerate}[label*=(\arabic*)]
		\item \textit{Preparation} 
		
    For each round, Alice chooses a pure state $\ket{\psi_j}_a$ with probability $p_j$, where $j \in \{ 0_Z, 1_Z, 0_X\}$, and sends it to Bob through the quantum channel. Emissions of $\ket{\psi_{0_X}}_a$ ($\ket{\psi_{0_Z}}_a$ and $\ket{\psi_{1_Z}}_a$) are considered to belong to the $X$ ($Z$) basis.

	\item \textit{Detection} 
		
    Bob measures the incoming signals in either the $Z$ or the $X$ basis, which he chooses with probabilities $p_Z$ and $p_X = 1 - p_Z$, respectively.
		
	\item \textit{Sifting} 
		
    Bob announces which rounds were detected, and Alice and Bob reveal their basis choices in those rounds. Let $\mathcal{K}_Z$ be the set of detected rounds in which both users employed the $Z$ basis, and let $\mathcal{T}_X$ be the set of detected rounds in which Bob employed the $X$ basis. Then,
    
		\begin{enumerate}[label=(\arabic{enumi}.\arabic*), ref=\arabic{enumi}.\arabic*] 
			\item Alice (Bob) defines her (his) sifted key as the bit values associated with her emissions (his measurement results) in the rounds in $\mathcal{K}_Z$.
			\item For all rounds in $\mathcal{T}_X$, Bob announces his measurement result. 
		\end{enumerate}

    \item \textit{Tag assignment}
    
    Alice probabilistically assigns a tag to all rounds in $\mathcal{T}_X$, depending on her choice of state and Bob's measurement result. Namely, if she chose the state $\ket{\psi_j}_a$ and Bob obtained measurement result $(\alpha \oplus 1)_{X}$, for $\alpha \in \{0,1\}$, she assigns a tag of $t_{\alpha} \in \{\rm pos_\alpha, neg_\alpha\}$ with probability $p_{t_{\alpha}|j,X_B}$, given by \cref{eq:tag_probability}. Then, she calculates $N^{(\alpha \oplus 1)_X}_{t_\alpha}$, the number of detected events with a tag of $t_{\alpha}$ in which Bob obtained measurement result $(\alpha \oplus 1)_X$.
	
	\item \textit{Parameter estimation}
    
    Alice uses the values of $\{N^{(\alpha \oplus 1)_X}_{t_\alpha}\}$ to obtain an upper bound $N_{\rm ph}^{\rm U}$ on $N_{\rm ph}$, the number of phase errors in her sifted key, using \cref{eq:Nph_pm}.

	\item \textit{Postprocessing} 
		
		\begin{enumerate}[label=(\arabic{enumi}.\arabic*), ref=\arabic{enumi}.\arabic*] 
			\item  \textit{Error correction}: Alice sends Bob a pre-fixed amount $\lambda_{\textrm{EC}}$ of syndrome information bits through an authenticated public channel, which Bob uses to correct errors in his sifted key.
			
			\item  \textit{Error verification}: Alice and Bob compute a hash of their error-corrected keys using a random universal hash function, and check whether they are equal. If so, they continue to the next step; otherwise, they abort the protocol.
			
			\item \textit{Privacy amplification:} Alice and Bob extract a secret key pair $(S_A, S_B)$ of length $\abs{S_A} = \abs{S_B} = \ell$ from their error-corrected keys using a random two-universal hash function.
		\end{enumerate}
		
	\end{enumerate}

	\section{Description of the MDI protocol}
	\label{app:protocol_description_mdi}
	\begin{enumerate}[label*=(\arabic*)]
		\item \textit{Quantum communication}
		
    For each round, Alice (Bob) selects the state $\ket{\psi_j}_a$ ($\ket{\psi_s'}_b$) with probability $p_j$ ($p_s'$), where $j$ ($s$) $\in \{ 0, 1, \tau\}$, and sends it to an untrusted middle node Charlie, who announces whether or not he obtained a successful projection to a Bell state. Emissions for which $j \in \{0,1\}$ ($s \in \{0,1\}$) are considered to belong to the $Z$ basis.     
    
	\item \textit{Sifting} 
	
	Alice and Bob announce their basis choices in the detected rounds. Then, they assign all detected rounds in which at least one of them used the $X$ basis to set $\mathcal{T}_{\rm d}$. Also, for each detected round in which both chose the $Z$ basis, they assign it to set $\mathcal{K}_{\rm d}$ with probability $p_{\mathcal{K}\vert\mathcal{Z}}$, or to set $\mathcal{T}_{\rm d}$ with probability $p_{\mathcal{T}\vert\mathcal{Z}} = 1 -p_{\mathcal{K}\vert\mathcal{Z}}$. 
	Then, they announce these assignments, and
		\begin{enumerate}[label=(\arabic{enumi}.\arabic*), ref=\arabic{enumi}.\arabic*] 
			\item Alice (Bob) defines her (his) sifted key as her (his) choices of $j$ ($s$) in the rounds in $\mathcal{K}_{\rm d}$.
			
			\item For all rounds in $\mathcal{T}_{\rm d}$, Alice and Bob announce their choice of $j$ and $s$.
		\end{enumerate}

    \item \textit{Tag assignment}
    
    Alice and Bob assign a tag $t\in\set{\textrm{pos},\textrm{neg}}$ to each round in $\mathcal{T}_{\rm d}$ with probability $p_{t|j,s,\mathcal{T}}$, give by \cref{eq:p_tjsT}. Then, they calculate $N_{t}$, the number of detected events with a tag of $t$.
	
	\item \textit{Parameter estimation}
    
    Alice and Bob substitute the values of $N_{\textrm{pos}}$ and $N_{\textrm{neg}}$ in \cref{eq:Nvir4MDI} to obtain an upper bound $N_{\rm ph}^{\rm U}$ on $N_{\rm ph}$, the number of errors in the sifted key.

	\item \textit{Postprocessing} 
		
		\begin{enumerate}[label=(\arabic{enumi}.\arabic*), ref=\arabic{enumi}.\arabic*] 
			\item  \textit{Error correction}: Alice sends Bob a pre-fixed amount $\lambda_{\textrm{EC}}$ of syndrome information bits through an authenticated public channel, which Bob uses to correct errors in his sifted key.
			
			\item  \textit{Error verification}: Alice and Bob compute a hash of their error-corrected keys using a random universal hash function, and check whether they are equal. If so, they continue to the next step; otherwise, they abort the protocol.
			
			\item \textit{Privacy amplification:} Alice and Bob extract a secret key pair $(S_A, S_B)$ of length $\abs{S_A} = \abs{S_B} = \ell$ from their error-corrected keys using a random two-universal hash function.
		\end{enumerate}
		
	\end{enumerate}

\section{Channel model for the MDI protocol}\label{app:channelModelMDI}

In this Appendix, we present the channel model used in our simulations of the MDI LT protocol, which is based on the single-photon version of the original MDI QKD scheme~\cite{lo2012measurement}. Specifically, we assume that Alice and Bob prepare polarised single-photon states in the form of \cref{eq:transmitted_states}, where here $\ket{0_Z}$ and $\ket{1_Z}$ denote the horizontally and vertically polarised single-photon states, respectively. After the preparation, Alice (Bob) sends the transmitted states to the intermediate party Charlie through a lossy quantum channel of transmittance $\eta_A$ ($\eta_B$), who interferes the two incoming signals in a 50:50 beamsplitter, which has on each output port a polarising beamsplitter (PBS) that separates the horizontal and vertical modes. Now, let $h_1$ and $v_1$ ($h_2$ and $v_2$) be the threshold detectors placed at horizontal and vertical output port of the first (second) PBS, respectively, and let $p_d$ be the dark-count probability of each detector. After the measurement, Charlie announces the Bell state $\Psi^+$ ($\Psi^-$) if he observes clicks in $h_1$ and $v_1$, or $h_2$ and $v_2$ ($h_1$ and $v_2$, or $h_1$ and $v_2$). Then, it is easy to prove that the conditional probability that Charles announces the Bell state $\Psi^{\pm}$ given that Alice and Bob selected the states $\ket{\psi_j}_a$ and $\ket{\psi_s}_b$, respectively, is
\begin{eqnarray}
\text{P}^{\Psi^\pm}_{j,s}&=&(1-p_d)^2\bigg[\frac{\eta_A\eta_B}{2}\sin^2(\kappa(\theta_j\pm\theta_s')) +p_d\frac{\eta_A\eta_B}{2}(1+\cos(2\kappa\theta_j)\cos(2\kappa\theta_s'))\\
&&+p_d(1-\eta_A)\eta_B+p_d\eta_A(1-\eta_B)+2p_d^2(1-\eta_a)(1-\eta_b)\bigg].\nonumber
\end{eqnarray}

\section{Alternative analysis using concentration inequalities for dependent random variables}
\label{app:analysis_dependent_inequalities}

In this Appendix, we present an alternative analysis that requires the application of a concentration inequality for sums of dependent Bernoulli random variables. This alternative analysis is a simplified version of that of Ref.~\cite{mizutani2}, which considers the emission of weak coherent pulses rather than single photons. In Ref.~\cite{mizutani2}, Azuma's inequality \cite{azuma} is the concentration inequality applied. Here, we also present a new security proof based on the application of the recently proposed Kato's inequality \cite{kato}.   First, we introduce the concentration inequalities that we consider, and then we provide the analysis. 

\subsection{Concentration inequalities }
Let $\xi_1,..., \xi_N$ be a sequence of Bernoulli random variables, and let $\Lambda_l = \sum_{u=1}^{l} \xi_u$. Let $\mathcal{F}_{l}$ be its natural filtration, i.e. the $\sigma$-algebra generated by $\{\xi_1,...,\xi_l\}$.
\subsubsection{Azuma's inequality}
    According to Azuma's inequality \cite{azuma},
    \begin{equation}
    \label{eq:azuma}
    \begin{gathered}
         \Pr \left[\Lambda_n - \sum_{u=1}^{N} \Pr(\xi_u = 1 \vert \mathcal{F}_{u-1}) \geq b \sqrt{N} \right] \leq \exp \left[-\frac{b^2}{2} \right],\\  
         \Pr \left[\sum_{u=1}^{N} \Pr(\xi_u = 1 \vert \mathcal{F}_{u-1}) - \Lambda_n \geq b \sqrt{N} \right] \leq \exp \left[-\frac{b^2}{2} \right].
    \end{gathered}
    \end{equation}
    Equating the right hand sides to to $\varepsilon_{\rm A}$ and solving for $b$, we have that
    \begin{equation}
    \label{eq:azuma_final}    
    \begin{gathered}
	\sum_{u=1}^{N} \Pr \left(\xi_u = 1 \vert\xi_1,..., \xi_{u-1}\right) \leq \Lambda_N  + \Delta_{\rm A}, \\
	\Lambda_N \leq  \sum_{u=1}^{N} \Pr \left(\xi_u = 1 \vert\xi_1,..., \xi_{u-1}\right) + \Delta_{\rm A},
	\end{gathered}
    \end{equation}
    except with probability at most $\varepsilon_{\rm A}$ for each of the bounds, where $\Delta_{\rm A} = \sqrt{2 N \ln \varepsilon_{\rm A}^{-1}}$.
    
    \subsubsection{Kato's inequality}

	According to Kato's inequality \cite{kato}, for any $n$, and any $a,b$ such that $b\geq \abs{a}$,
        \begin{equation}
    \label{eq:katobound_upper}
        \Pr \left[\sum_{u=1}^{N} \Pr(\xi_u = 1 \vert \mathcal{F}_{u-1}) - \Lambda_N  \geq \left[b+a \left(\frac{2 \Lambda_N}{N} - 1 \right) \right] \sqrt{N} \right] \leq \exp \left[\frac{-2(b^2-a^2)}{(1+\frac{4a}{3 \sqrt{N}})^2}  \right].
    \end{equation}
    By replacing $\xi_l \to 1-\xi_l$ and $a \to -a$ in \cref{eq:katobound_upper}, we also derive \cite{curraslorenzo2019tight}
     \begin{equation}
        \label{eq:katobound_lower}
        \Pr \left[\Lambda_N - \sum_{u=1}^{N} \Pr(\xi_u = 1 \vert \mathcal{F}_{u-1})  \geq \left[b+a \left(\frac{2 \Lambda_N}{N} - 1 \right) \right] \sqrt{N} \right] \leq \exp \left[\frac{-2(b^2-a^2)}{(1-\frac{4a}{3 \sqrt{N}})^2} \right].
    \end{equation}  
    By isolating $\Lambda_N$ in \cref{eq:katobound_lower}, we derive,
    \begin{equation}
            \label{eq:katobound_inverse}
            \Pr \left[\Lambda_N \geq \frac{N}{\sqrt{N}-2a} \left(\frac{1}{\sqrt{N}}\sum_{u=1}^{N} \Pr(\xi_u = 1 \vert \mathcal{F}_{u-1}) + b -a  \right)\right] \leq \exp \left[\frac{-2(b^2-a^2)}{(1-\frac{4a}{3 \sqrt{N}})^2} \right],
    \end{equation}
    which holds when $a \leq \sqrt{N}/2$.
    
    In the following, we will use \cref{eq:katobound_upper} to obtain an upper bound on $\sum_{u=1}^{N} \Pr(\xi_u = 1 \vert \mathcal{F}_{u-1})$, \cref{eq:katobound_lower} to obtain a lower bound on $\sum_{u=1}^{N} \Pr(\xi_u = 1 \vert \mathcal{F}_{u-1})$, and \cref{eq:katobound_inverse} to obtain an upper bound on $\Lambda_N$.
    
    \subsubsection*{Upper bound on the sum of probabilities}
    
    Before running the protocol, one should use previous knowledge of the channel to come up with a prediction $\tilde{\Lambda}_N$ of the value of  $\Lambda_N$ that one expects to obtain. Then, one calculates the values of $a$ and $b$ that would minimise the deviation term in \cref{eq:katobound_upper} if the realisation of $\Lambda_N$ equalled $\tilde{\Lambda}_N$, for a fixed failure probability $\varepsilon_{\rm K}$. These are the solution of the optimisation problem
        \begin{equation}
    \begin{aligned}
    &\min_{a,b} &  \left[b+a \left(\frac{2 \tilde{\Lambda}_N}{N} - 1 \right) \right] \sqrt{N} \\
    &\textrm{s.t. }  &\exp \left[\frac{-2(b^2-a^2)}{(1+\frac{4a}{3 \sqrt{N}})^2} \right] = \varepsilon_{\rm K}, \\
    && b \geq \abs{a},
    \end{aligned}
    \end{equation}
    which can be expressed as
    \begin{equation}
    \begin{gathered}
        a = \frac{3 \left(72 \sqrt{n} \tilde{\Lambda}_N  (n-\tilde{\Lambda}_N)\ln \varepsilon_{\rm K}-16 N^{3/2} \ln^2\varepsilon_{\rm K}+9 \sqrt{2} (N-2 \tilde{\Lambda}_N) \sqrt{-N^2 \ln \varepsilon_{\rm K}  (9 \tilde{\Lambda}_N (n-\tilde{\Lambda}_N)-2 N \ln \varepsilon_{\rm K})}\right)}{4 (9 N-8\ln
	\varepsilon_{\rm K}) (9 \tilde{\Lambda}_N (n-\tilde{\Lambda}_N)-2 N \ln\varepsilon_{\rm K})}, \\
b = \frac{\sqrt{18 a^2 N-\left(16 a^2+24 a \sqrt{n}+9 N\right) \ln\varepsilon_{\rm K}}}{3 \sqrt{2N}}.
   \end{gathered}
    \end{equation}
    Then, we have that
	\begin{equation}
	\label{eq:katobound_upper_final}
		\sum_{u=1}^{N} \Pr \left(\xi_u = 1 \vert\xi_1,..., \xi_{u-1}\right) \leq \Lambda_N  + \Delta_K^{\rm U},
	\end{equation}
	except with probability $\varepsilon_{\rm K}$, where
	\begin{equation}
	\Delta_K^{\rm U} = \left[b+a \left(\frac{2 \Lambda_N}{N} - 1 \right) \right] \sqrt{N}.
	\end{equation}

    \subsubsection*{Lower bound on the sum of probabilities}
  
      Similarly to the previous case, one should use previous knowledge of the channel to come up with a prediction $\tilde{\Lambda}_N$ of the value of  $\Lambda_N$ that one expects to obtain after running the protocol. Then, one calculates the values of $a$ and $b$ that would minimise the deviation term in \cref{eq:katobound_lower} if the realisation of $\Lambda_N$ equalled $\tilde{\Lambda}_N$, for a fixed failure probability $\varepsilon_{\rm K}$. These are the solution of the optimisation problem
        \begin{equation}
    \begin{aligned}
    &\min_{a,b} &  \left[b+a \left(\frac{2 \tilde{\Lambda}_N}{N} - 1 \right) \right] \sqrt{N} \\
    &\textrm{s.t. }  &\exp \left[\frac{-2(b^2-a^2)}{(1-\frac{4a}{3 \sqrt{N}})^2} \right] = \varepsilon_{\rm K}, \\
    && b \geq \abs{a},
    \end{aligned}
    \end{equation}
    which can be expressed as
    \begin{equation}
    \begin{gathered}
        a = \frac{3 \left(-72 \sqrt{N} \tilde{\Lambda}_N  (n-\tilde{\Lambda}_N)\ln \varepsilon_{\rm K}+16 N^{3/2} \ln^2\varepsilon_{\rm K}+9 \sqrt{2} (N-2 \tilde{\Lambda}_N) \sqrt{-N^2 \ln \varepsilon_{\rm K}  (9 \tilde{\Lambda}_N (n-\tilde{\Lambda}_N)-2 N \ln \varepsilon_{\rm K})}\right)}{4 (9 N-8\ln
	\varepsilon_{\rm K}) (9 \tilde{\Lambda}_N (n-\tilde{\Lambda}_N)-2 N \ln\varepsilon_{\rm K})}, \\
b = \frac{\sqrt{18 a^2 N-\left(16 a^2-24 a \sqrt{n}+9 N\right) \ln\varepsilon_{\rm K}}}{3 \sqrt{2N}}.
   \end{gathered}
    \end{equation}
    Then, we have that
	\begin{equation}
	\label{eq:katobound_lower_final}
	\sum_{u=1}^{N} \Pr \left(\xi_u = 1 \vert \mathcal{F}_{u-1}\right) \geq \Lambda_N  - \Delta_K^{\rm L},
	\end{equation}
	except with probability $\varepsilon_{\rm K}$, where
	\begin{equation}
	\Delta_K^{\rm L} = \left[b+a \left(\frac{2 \Lambda_N}{N} - 1 \right) \right] \sqrt{N}.
	\end{equation}
  
    \subsubsection*{Upper bound on the actual value}
     In this case, we assume that we have an upper bound $S_N$ on the sum of probabilities $\sum_{u=1}^{N} \Pr \left(\xi_u = 1 \vert \mathcal{F}_{u-1}\right)$, and we want to obtain an upper bound on $\Lambda_N$. Before running the protocol one should use previous knowledge to come up with a prediction $\tilde{S}_N$ of the  value of the upper bound $S_N$ that one expects to obtain. Then, one calculates the values of $a$ and $b$ that would minimise the deviation term in \cref{eq:katobound_inverse} if the prediction comes true. These are the solution of the
     optimisation problem
   \begin{equation}
    \begin{aligned}
    &\min_{a,b} & \frac{N}{\sqrt{N}-2a} \left(\frac{1}{\sqrt{N}}\tilde{S}_N + b -a  \right) \\
    &\textrm{s.t. }  &\exp \left[\frac{-2(b^2-a^2)}{(1-\frac{4a}{3 \sqrt{N}})^2} \right] = \varepsilon_{\rm K}, \\
    && b \geq \abs{a},
    \end{aligned}
    \end{equation}
    whose analytical solution is
    \begin{equation}
    \begin{gathered}
    a = \frac{3 \sqrt{N} \left(9 \ln \varepsilon_{\rm K} \left(3 N^2-8 N \tilde{S}_N+8 \tilde{S}_N^2\right)+9 (N-2 \tilde{S}_N) \sqrt{N \ln \varepsilon_{\rm K}  (N \ln \varepsilon_{\rm K} +18 \tilde{S}_N (\tilde{S}_N-N))}+4 n \ln ^2(\varepsilon_{\rm K} )\right)}{4
   \left(36 \ln \varepsilon_{\rm K}  \left(N^2-2 N \tilde{S}_N+2 \tilde{S}_N^2\right)+4 N \ln ^2\varepsilon_{\rm K} +81 N \tilde{S}_N (N-\tilde{S}_N)\right)}, \\
   b = \frac{\sqrt{18 a^2 N-\left(16 a^2-24 a \sqrt{N}+9 N\right) \ln\varepsilon_a}}{3 \sqrt{2N}}.
    \end{gathered}
    \end{equation}
    Then, we have that,
    \begin{equation}
    \label{eq:katobound_inverse_final}
        \Lambda_N \geq \frac{N}{\sqrt{N}-2a} \left(\frac{1}{\sqrt{N}} S_N + b -a  \right),
    \end{equation}
    except with probability $\varepsilon_{\rm K}$.

\subsection{Analysis}
We assume a virtual protocol in which Alice prepares $N_{\rm tot}$ copies of the entangled state in \cref{eq:vir_prot}, and sends all subsystems $B$ to Bob through the untrusted quantum channel. Then, Bob performs a quantum non-demolition measurement on each system $B$, learning which rounds produce a click on his detectors, and saving the system $B$ of these detected rounds in a quantum memory. Let $N$ be the number of detected rounds. For each detected round $u = \{1,2,\ldots,N\}$, Alice measures her ancilla $S$, and Bob measures $B$ in the $X$ basis; except if Alice obtained $S = 5$, in which case Bob measures $B$ in the $Z$ basis. Let $\xi_u = (i, j)$ represent the event "Alice learns that she emitted $i$ and Bob obtains measurement result $j$". More specifically, Alice learns that she emitted $i=\{{\rm vir_0}, {\rm vir_1}, 0_Z, 1_Z, 0_X\}$ if she obtained $S = \{0,1,2,3,4\}$ in her measurement of system $S$, respectively. Events in which she obtained $S = 5$ are ignored in the analysis. Then, using the fact that the virtual states can be written as an operator-form linear function of the actual states as in \cref{eq:qubit_space}, one can show that
\begin{equation}
\label{eq:avg_ph_error}
    \sum_{u = 1}^{N} \Pr[\xi_u = ({\rm vir_\alpha},\, \alpha \oplus 1 ) \vert \mathcal{F}_{u-1}] = \sum_{i = \{0_Z, 1_Z, 0_X\}} \frac{p_{\rm vir_\alpha} p_{Z_B} c_{i \vert {\rm vir_\alpha}}}{p_i p_{X_B}} \sum_{u = 1}^{N} \Pr[\xi_u = (i,\, \alpha \oplus 1 )\vert \mathcal{F}_{u-1}],
\end{equation}
where $\mathcal{F}_{u-1}$ is the $\sigma$-algebra generated by $\{\xi_1,...,\xi_{u-1}\}$. Now one needs to apply a concentration bound for sums of dependent random variables to substitute the sums of probabilities in \cref{eq:avg_ph_error} by the actual values.

\subsubsection{Using Azuma's inequality}
Applying \cref{eq:azuma_final} to \cref{eq:avg_ph_error}, we have that
\begin{equation}
N_{\rm vir_\alpha}^{\alpha \oplus 1} \leq \sum_{i = \{0_Z, 1_Z, 0_X\}} \frac{p_{\rm vir_\alpha} p_{Z_B} c_{i \vert {\rm vir_\alpha}}}{p_i p_{X_B}} (N_{i}^{\alpha \oplus 1} + \delta_i) + \Delta_{\rm A} := \overline{N}_{\rm vir_\alpha}^{\alpha \oplus 1},
\end{equation}
except with probability $4 \varepsilon_{\rm A}$, where $\varepsilon_{\rm A}$ is the failure probability of each aplication of Azuma's inequality, which has been applied four times; and $\delta_i = \Delta_{\rm A}$ ($\delta_i = -\Delta_{\rm A}$) if $c_{i \vert {\rm vir_\alpha}}$ is positive (negative).
Then, the number of phase errors is upper bounded by
\begin{equation}
    N_{\rm ph} \leq  \overline{N}_{\rm vir_0}^{1} + \overline{N}_{\rm vir_1}^{0},
\end{equation}
except with probability $\varepsilon = 8 \varepsilon_{\rm A}$.

Using a similar analysis, for the MDI protocol, we have that
\begin{equation}
N_{\rm ph} \leq \sum_{j,s} \frac{p_{\rm ph} c_{j,s}}{p_{j,s,\mathcal{T}}} (N_{j,s,\mathcal{T}} + \delta_{j,s}) + \Delta_{\rm A}
\end{equation}
except with probability $\varepsilon = 9\varepsilon_{\rm A}$, where $N_{j,s,\mathcal{T}}$ is the number of detected test rounds in which the user emitted $\ket{\psi_{j,s}}_{a,b}$, and $\delta_{j,s} = \Delta_{\rm A}$ ($\delta_{j,s} = - \Delta_{\rm A}$) if $c_{j,s}$ is positive (negative).

\subsubsection{Using Kato's inequality}
Applying \cref{eq:katobound_upper_final} and \cref{eq:katobound_lower_final} to \cref{eq:avg_ph_error}, we have that
\begin{equation}
\sum_{u = 1}^{N} \Pr[\xi_u = ({\rm vir_\alpha},\, \alpha \oplus 1 ) \vert \mathcal{F}_{u-1}] \leq \sum_{i = \{0_Z, 1_Z, 0_X\}} \frac{p_{\rm vir_\alpha} p_{Z_B} c_{i \vert {\rm vir_\alpha}}}{p_i p_{X_B}} (N_{i}^{\alpha \oplus 1} + \delta_i):= S_{\rm vir_\alpha},
\end{equation}
except with probability $3\varepsilon_{\rm K}$, where $\delta_i = \Delta_{\rm K}^{\rm U}$ ($\delta_i = -\Delta_{\rm K}^{\rm L}$) if $c_{i \vert {\rm vir_\alpha}}$ is positive (negative). Substituting $S_n \to S_{\rm vir_\alpha}$ and $\Lambda_n \to N_{\rm vir_\alpha}^{\alpha \oplus 1}$ in \cref{eq:katobound_inverse_final}, we obtain an upper bound $\overline{N}_{\rm vir_\alpha}^{\alpha \oplus 1}$ which fails with probability $4 \varepsilon_{\rm K}$. Then, the number of phase errors is upper bounded by
\begin{equation}
    N_{\rm ph} \leq  \overline{N}_{\rm vir_0}^{1} + \overline{N}_{\rm vir_1}^{0},
\end{equation}
except with probability $\varepsilon = 8 \varepsilon_{\rm K}$.

Similarly, for the MDI protocol, we have that
\begin{equation}
\sum_{u = 1}^{N} \Pr[\xi_u = {\rm ph}  \vert \mathcal{F}_{u-1}] \leq \sum_{j,s} \frac{p_{\rm ph} c_{j,s}}{p_{j,s,\mathcal{T}}} (N_{j,s,\mathcal{T}} + \delta_{j,s}) := S_{\rm ph}
\end{equation}
except with probability $\varepsilon = 8\varepsilon_{\rm A}$, where $\delta_{j,s} = \Delta_{\rm K}^{\rm U}$ ($\delta_{j,s} = - \Delta_{\rm K}^{\rm L}$) if $c_{j,s}$ is positive (negative). Then, substituting $S_n \to S_{\rm ph}$ and $\Lambda_n \to N_{\rm ph}$ in \cref{eq:katobound_inverse_final}, we obtain an upper bound on $N_{\rm ph}$ which fails with probability $9 \varepsilon_{\rm K}$.
\end{document}